\documentclass {aastex631}

\newcommand\SDSSgalaxy{SDSS J025810.28-085719.2}
\newcommand\SDSSwebsite{\href{http://skyserver.sdss.org/dr17/VisualTools/quickobj?name=SDSS\%20J025810.28-085719.2\&ra=044.5428158997300\&dec=-08.9553581928100}{skyserver.sdss.org}}

\newcommand\CaIRfull{Ca II NIR triplet $\lambda\lambda$~8498,~8542,~8662~\AA}
\newcommand\HeIfull{He I $\lambda$~10830~\AA}

\newcommand\flamunit{erg/s/cm$^2$/\AA}
\newcommand\flamunitintegrate{erg/s/cm$^2$}
\newcommand\Izzoinprep{Izzo et al. (in prep.)}
\newcommand\Patricia{Patricia Schady}
\newcommand\Medlerinprep{Medler et al. (in prep.)}

\graphicspath{{./}{Figures/}}
\begin{document}

\title{The Redshift of GRB~190829A/SN~2019oyw: A Case Study of GRB-SN Evolution.}

\correspondingauthor{Kornpob Bhirombhakdi}
\email{bkornpob@gmail.com}

\author[0000-0003-0136-1281]{Kornpob Bhirombhakdi}
\affiliation{Space Telescope Science Institute, 3700 San Martin Dr, Baltimore, MD 21218, USA}

\author[0000-0002-6652-9279]{Andrew S. Fruchter}
\affiliation{Space Telescope Science Institute, 3700 San Martin Dr, Baltimore, MD 21218, USA}

\author{Andrew J. Levan}
\affiliation{Department of Astrophysics/IMAPP, Radboud University Nijmegen, P.O. Box 9010, 6500 GL Nijmegen, The Netherlands}
\affiliation{Department of Physics, University of Warwick, Coventry, CV4 7AL, UK}

\author[0000-0001-8646-4858]{Elena Pian}
\affiliation{INAF, Astrophysics and Space Science Observatory, Via P. Gobetti 101, 40129 Bologna, Italy}

\author{Paolo Mazzali}
\affiliation{Astrophysics Research Institute, Liverpool John Moores University, 146 Brownlow Hill, Liverpool L3 5RF, UK}
\affiliation{Max-Planck Institut f\"ur Astrophysik, Karl-Schwarzschild-Str. 1, D-85748 Garching, Germany}

\author[0000-0001-9695-8472]{Luca Izzo}
\affiliation{DARK, Niels Bohr Institute, University of Copenhagen, Copenhagen N, Denmark}
\affiliation{INAF - Osservatorio Astronomico di Capodimonte, Naples, Italy}

\author[0000-0002-5477-0217]{Tuomas Kangas}
\affiliation{Finnish Centre for Astronomy with ESO (FINCA), FI-20014 University of Turku, Finland}
\affiliation{Department of Physics and Astronomy, University of Turku, Vesilinnantie 5, FI-20500, Finland}

\author[0000-0002-3256-0016]{Stefano Benetti}
\affiliation{INAF – Osservatorio Astronomico di Padova, vicolo dell’Osservatorio 5, Padova I-35122, Italy}

\author[0000-0001-7186-105X]{Kyle Medler}
\affiliation{Astrophysics Research Institute, Liverpool John Moores University, 146 Brownlow Hill, Liverpool L3 5RF, UK}

\author[0000-0003-3274-6336]{Nial Tanvir}
\affiliation{School of Physics and Astronomy, University of Leicester, University Road, Leicester LE1 7RJ, UK}

%% Note that the \and command from previous versions of AASTeX is now
%% depreciated in this version as it is no longer necessary. AASTeX 
%% automatically takes care of all commas and "and"s between authors names.

%% AASTeX 6.31 has the new \collaboration and \nocollaboration commands to
%% provide the collaboration status of a group of authors. These commands 
%% can be used either before or after the list of corresponding authors. The
%% argument for \collaboration is the collaboration identifier. Authors are
%% encouraged to surround collaboration identifiers with ()s. The 
%% \nocollaboration command takes no argument and exists to indicate that
%% the nearby authors are not part of surrounding collaborations.

%% Mark off the abstract in the ``abstract'' environment. 

\begin{abstract}
The nearby long gamma-ray burst (GRB) 190829A was observed using the $HST$/WFC3/IR grisms about four weeks to 500 days after the burst. We find the spectral features of its associated supernova, SN~2019oyw, are redshifted by several thousands km/s compared to the redshift of the large spiral galaxy on which it is superposed. This velocity offset is seen in several features but most clearly in \CaIRfull~(CaIR3). We also analyze VLT/FORS and X-shooter spectra of the SN and find strong evolution with time of its P-Cygni features of CaIR3 from the blue to the red. However, comparison with a large sample of Type Ic-BL and Ic SNe shows no other object with the CaIR3 line as red as that of SN~2019oyw were it at the $z = 0.0785$ redshift of the disk galaxy. This implies that SN 2019oyw is either a highly unusual SN or is moving rapidly with respect to its apparent host. Indeed, using CaIR3 we find the redshift of SN~2019oyw  is  $0.0944 \leq z \leq 0.1156$. The GRB-SN is superposed on a particularly dusty region of the massive spiral galaxy; therefore, while we see no sign of a small host galaxy behind the spiral, it could be obscured. Our work provides a surprising result on the origins of GRB 190829A, as well as insights into the time evolution of GRB-SNe spectra and a method for directly determining the redshift of a GRB-SN using the evolution of strong spectral features such as CaIR3.
\end{abstract}

%% Keywords should appear after the \end{abstract} command. 
%% The AAS Journals now uses Unified Astronomy Thesaurus concepts:
%% https://astrothesaurus.org
%% You will be asked to selected these concepts during the submission process
%% but this old "keyword" functionality is maintained in case authors want
%% to include these concepts in their preprints.

%% From the front matter, we move on to the body of the paper.
%% Sections are demarcated by \section and \subsection, respectively.
%% Observe the use of the LaTeX \label
%% command after the \subsection to give a symbolic KEY to the
%% subsection for cross-referencing in a \ref command.
%% You can use LaTeX's \ref and \label commands to keep track of
%% cross-references to sections, equations, tables, and figures.
%% That way, if you change the order of any elements, LaTeX will
%% automatically renumber them.
%%
%% We recommend that authors also use the natbib \citep
%% and \citet commands to identify citations.  The citations are
%% tied to the reference list via symbolic KEYs. The KEY corresponds
%% to the KEY in the \bibitem in the reference list below. 

\section{Introduction} \label{sec:intro}

The long-duration gamma-ray burst~(GRB)~190829A \citep{deUgarte2019,Lipunov2019_GCN_SN2019oyw,Volnova2019_GCN_SN2019oyw} was detected by the {\it Fermi}/Gamma-ray Burst Monitor (GBM) on 29$^{th}$~August~2019, 19:55:53.13 UTC (henceforth, we adopted this as $T_0$ for the event; \cite{Fermi2019}), and shortly after by the {\it Swift}/Burst Alert Telescope \citep{Dichiara2019_GCN25552}. Its afterglow was detected in multiple wavelengths from X-ray to radio \citep{Chand2020,Dichiara2021,Fraija2020,Hu2021,Salafia2021}, and revealed its location to be superposed on the galaxy \SDSSgalaxy~(henceforth, we referred to this galaxy in short as the ``SDSS galaxy''; for further information see \SDSSwebsite; \cite{SDSS_DR17_2022}; also Figure \ref{fig:galfull}).\footnote{This SDSS galaxy is also known as 2MASX~J02581029-0857189 \citep{2MASS_2006}.} Very high energy (VHE) photons in the tera-electronvolt range were detected from the event \citep{deNaurois2019,HESS2021}, while the search for associated neutrinos yielded only an upper limit \citep{ANTARES2021}. The detection of Ca II H/K $\lambda\lambda$ 3934, 3969 \AA~absorption in the afterglow \citep{Hu2021,Valeev2019} gave the redshift estimate $z = 0.0785$, which is consistent with the observed redshift of that galaxy. This bright spiral has a stellar mass of order of 10$^{12}$ M$_\odot$ \citep{Gupts2022_GRB190829A_host}, the largest stellar mass of any proposed GRB host. The galaxy also exhibits characteristics of an active galactic nucleus (AGN; \Izzoinprep; \Patricia~via personal communication), which would be the first for a long GRB host. 
Moreover, the sight-line to the GRB was estimated to have $A_V = 2.33$ mag \citep{ZhangLuLu2021_GRB190829A+environment}, making it unusually dusty for a long GRB environment at a redshift $z< 1$. Despite being atypical for a long GRB host \citep{Fruchter2006_LGRB_CCSN_host,Japelj2018_IcBL_host,Kelly2014_SNIcBL_host,Modjaz2020_Ic_IcBL_host,Perley2016_grbHost_review}, the SDSS galaxy seems to be the obvious choice considering the location of the burst and the fact that all observed absorption features in the GRB spectrum appear to come from this galaxy.

Perhaps in part because it was so nearby, GRB~190829A itself was found to have a number of somewhat unusual properties. For example, the Amati correlation, which typically characterizes long GRBs \citep{Amati2006_Eiso-Ep_correlation}, was violated by one, but not another of two main emission events from this burst \citep{Chand2020}. Although it was a VHE burst displaying photons with energy above 100 GeV, its peak photon energy was among the lowest of known VHE-GRBs (see \cite{Sato2022_VHE_GRB_SynchrotronSelfCompton} and references therein). It also had the smallest isotropic energy when compared to other VHE-GRBs \citep{Abdalla2019,Blanch2020_GRB201216C_VHE,Duan2019_VHE-GRB180720B,MAGIC2019,Ravasio2019_VHE-GRB190114C}. 

However, it is the supernova associated with the GRB~190829A, SN~2019oyw, which most interests us in this paper. SN~2019oyw emerged a few days after the GRB trigger \citep{deUgarte2019,Lipunov2019_GCN_SN2019oyw,Volnova2019_GCN_SN2019oyw}, and reached its $i$-band peak by $\sim$20 days \citep{Hu2021}. 
% This was the first confirmed supernova associated with a VHE-GRB. 
Its early spectra showed characteristic features of a type Ic-BL \citep{Fraija2020,Hu2021}, similar to other GRB-SNe \citep{Cano2014_ThreeGRB-SNe_2013ez_slowest,Modjaz2016_spectra_Ic_IcBL_GRB-SNe} and in particular the prototype GRB-SN~1998bw \citep{Patat2001_SN1998bw}. Broad absorption features of Si II $\lambda$ 6355 $\AA$ and the Ca II near-infrared (NIR) triplet $\lambda\lambda$ 8498, 8542, 8662 $\AA$ (henceforth CaIR3, using the terminology of \cite{Silverman2015}) were identified \citep{Hu2021}. 

In this paper (Section \ref{sec:data}), we present late-time observations of SN~2019oyw from $HST$/WFC3 observed between 2019-09-28 and 2021-01-09 (i.e., 30 -- 499 days after the trigger in the observing frame). Optical and NIR images and NIR grism data were obtained. Our analysis of these data (Section \ref{sec:analysis}) shows that the spectral features associated with the supernova at late time ($\sim$50 days) are redshifted by several thousands of km/s compared to the redshift of the large spiral on which it is superposed. With additional data from the Very Large Telescope (VLT) provided by \citep{Medler_190829Athesis_ljmu21959} and \Izzoinprep, we analyze spectral features of the supernova together with a sample of GRB-SNe, SNe Ic-BL (without associated GRB), and SNe Ic, and show that SN~2019oyw is the only outlier (Section \ref{subsec:uncertainties}). We re-estimate the redshift that improves the feature alignment, and interpret this as surprising evidence that the SN progenitor exploded in or near the SDSS galaxy with a great peculiar velocity, or it was hosted by another galaxy of the order 100 Mpc behind the SDSS galaxy (Section \ref{subsec:interpret}). However, we find no evidence of another more distant host in the spectra or images, while the non-detection still does not rule out the possibility. Lastly, we summarize our results in Section \ref{sec:sum}.\footnote{We assume a $\Lambda$CDM cosmology with $H_0 = 70$ km/s/Mpc, $\Omega_m = 0.3$, $\Omega_{\Lambda} = 0.7$.}

\section{Observations} \label{sec:data}

\begin{deluxetable*}{cclhllc}
\tablenum{1}
\tablecaption{Summary of datasets from $HST$/WFC3 \label{table:hst}}
\tablewidth{0pt}
\tablehead{
\colhead{Proposal ID} & \colhead{PI} & \colhead{Epoch} & \nocolhead{MJD} & \colhead{Channel} & \colhead{Filter} &
\colhead{Total Exposure (s)}
}
\decimalcolnumbers
\startdata
15089 & E. Troja & 2019-09-28 & - & IR & F110W$^*$ & 148 \\
{} & {} & {} & {} & {} & F160W$^*$ & 148 \\
{} & {} & {} & {} & {} & G102 & 2012 \\
{} & {} & {} & {} & {} & G141 & 2012 \\
{} & {} & 2019-10-26 & - & IR & F110W$^*$ & 148 \\
{} & {} & {} & {} & {} & F160W$^*$ & 148 \\
{} & {} & {} & {} & {} & G102 & 2012 \\
{} & {} & {} & {} & {} & G141 & 2012 \\
15510 & N. Tanvir & 2019-11-25 & - & UVIS & F606W & 870 \\
{} & {} & {} & {} & {} & F814W & 870 \\
{} & {} & 2019-11-29 & {} & IR & F105W$^*$ & 349 \\
{} & {} & {} & {} & {} & F125W$^*$ & 349 \\
{} & {} & {} & {} & {} & F140W$^*$ & 698 \\
{} & {} & {} & {} & {} & G102 & 4095 \\
{} & {} & {} & {} & {} & G141 & 4095 \\
{} & {} & 2020-01-12 & - & UVIS & F606W & 900 \\
{} & {} & {} & {} & {} & F814W & 900 \\
{} & {} & {} & {} & IR & F105W$^*$ & 349 \\
{} & {} & {} & {} & {} & F125W$^*$ & 349 \\
{} & {} & {} & {} & {} & F140W$^*$ & 349 \\
{} & {} & {} & {} & {} & G102 & 4095 \\
{} & {} & {} & {} & {} & G141 & 4095 \\
16042 & A. Levan & 2020-02-11 & - & IR & F125W$^*$ & 698 \\
{} & {} & {} & {} & {} & F140W & 1198 \\
{} & {} & {} & {} & {} & G102 & 4095 \\
{} & {} & {} & {} & {} & G141 & 4095 \\
{} & {} & 2020-02-16 & - & UVIS & F606W & 1044 \\
{} & {} & 2020-06-24 & - & UVIS & F606W & 1044 \\
{} & {} & {} & {} & IR & F140W & 1198 \\
{} & {} & 2020-08-21 & - & UVIS & F606W & 1044 \\
{} & {} & {} & {} & IR & F140W & 1198 \\
16320 & A. Levan & 2021-01-09 & - & UVIS & F606W & 1044 \\
{} & {} & {} & {} & IR & F125W$^*$ & 698 \\
{} & {} & {} & {} & {} & F140W$^*$ & 1746 \\
{} & {} & {} & {} & {} & G102 & 4095 \\
{} & {} & {} & {} & {} & G141 & 4095 \\
\enddata
\tablecomments{a) $^*$ indicates that images taken with grism observations were included. See \cite{WFC3_Instrument_Handbook} for details. b) All of the data presented in this table were obtained from the MAST at the Space Telescope Science Institute. The specific observations analyzed can be accessed via \dataset[doi:10.17909/yz9m-xb66]{https://doi.org/10.17909/yz9m-xb66}.}
\end{deluxetable*}

GRB~190829A/SN~2019oyw was observed at several epochs between 2019-09-28 and 2021-01-09 (i.e., 30 -- 499 days after the trigger in the observing frame) by the Wide Field Camera 3 of the \textit{Hubble Space Telescope}~($HST$/WFC3; \cite{WFC3_Instrument_Handbook}). Both images and spectra were obtained in optical and NIR, as summarized in Table \ref{table:hst}. In this section, we briefly describe the data and the reduction process for photometry and spectroscopy from the $HST$, in Subsections \ref{subsec:reduction-phot} and \ref{subsec:reduction-spec}  respectively. In addition to the $HST$ data, we were provided additional data from the Very Large Telescope (VLT) by \cite{Medler_190829Athesis_ljmu21959} and \Izzoinprep. We briefly describe the VLT observations in Subsection \ref{subsec:data-vlt}. 

\subsection{$HST$/WFC3 Photometry} \label{subsec:reduction-phot}

\begin{figure}
\plotone{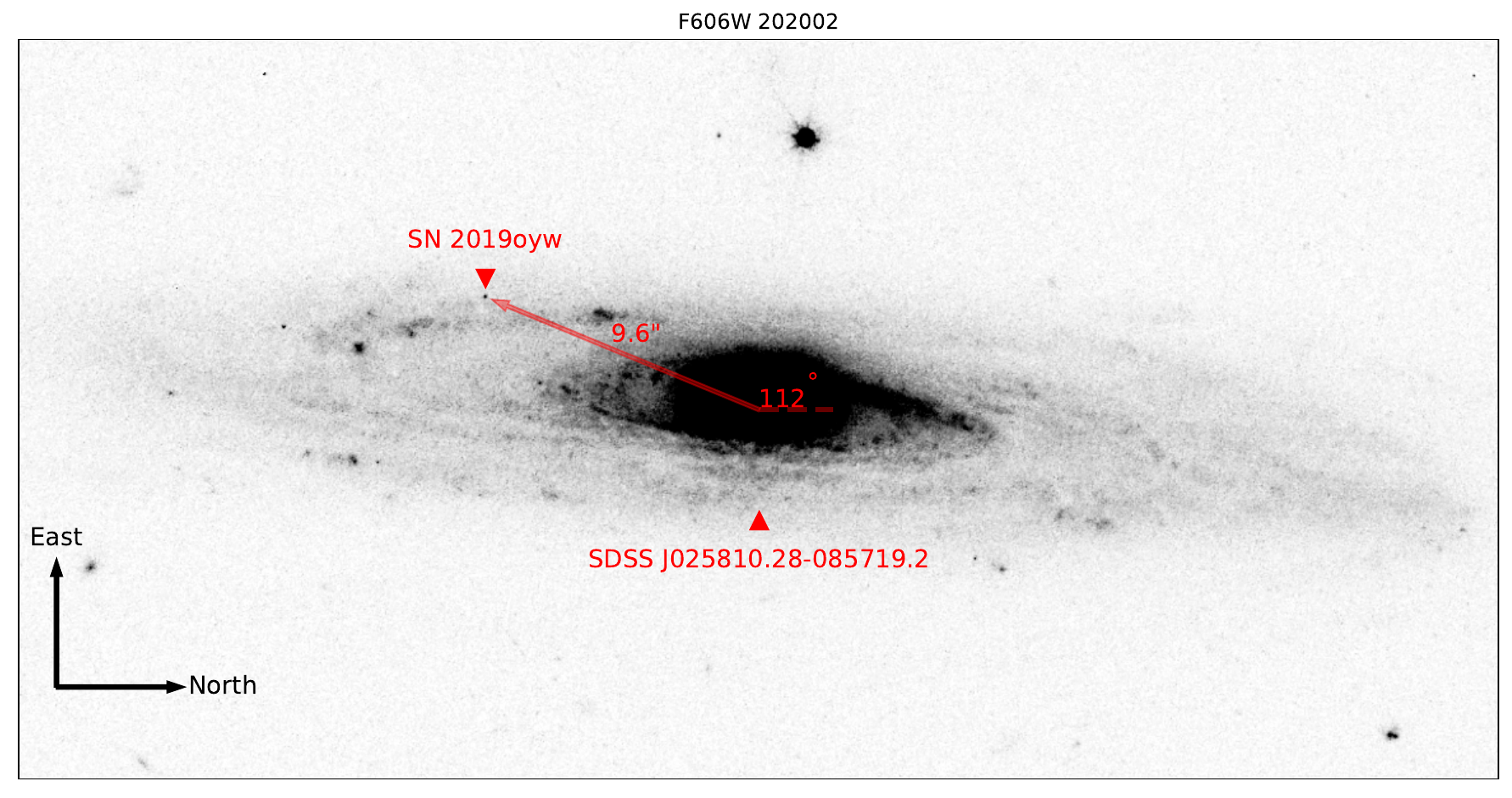}
\caption{Image of SN~2019oyw superposing on the outer disc of SDSS J025810.28-085719.2. This image was taken with $HST$/WFC3/UVIS/F606W on 2020-02-16. SN~2019oyw was $\sim$10$\arcsec$ from the brightest pixel at the galactic center in the direction $\sim$112\textdegree~with respect to north towards southeast. \label{fig:galfull}}
\end{figure}

Images of SN~2019oyw were taken by $HST$/WFC3 \citep{WFC3_Instrument_Handbook} in optical (UVIS channel) and NIR (IR channel). Table \ref{table:hst} summarizes the data. We followed the reduction process recommended by \cite{WFC3_Data_Handbook}. The data were pre-processed and downloaded from the Barbara A. Mikulski Archive for Space Telescopes (MAST; \url{https://archive.stsci.edu/}) in FLT/FLC format. The dithered images were combined using Drizzle \citep{Fruchter2002} as implemented in \textsc{AstroDrizzle} (Gonzaga).\footnote{\url{https://www.stsci.edu/scientific-community/software/drizzlepac.html}} Figure \ref{fig:galfull} shows an example of the processed image from F606W on 2020-02-16. The image clearly shows SN~2019oyw superposed on the outer disc of the \SDSSgalaxy~\citep{Dichiara2019_GCN25552,Hu2021,Oates2019_GCN25570}.

\begin{figure}
\plotone{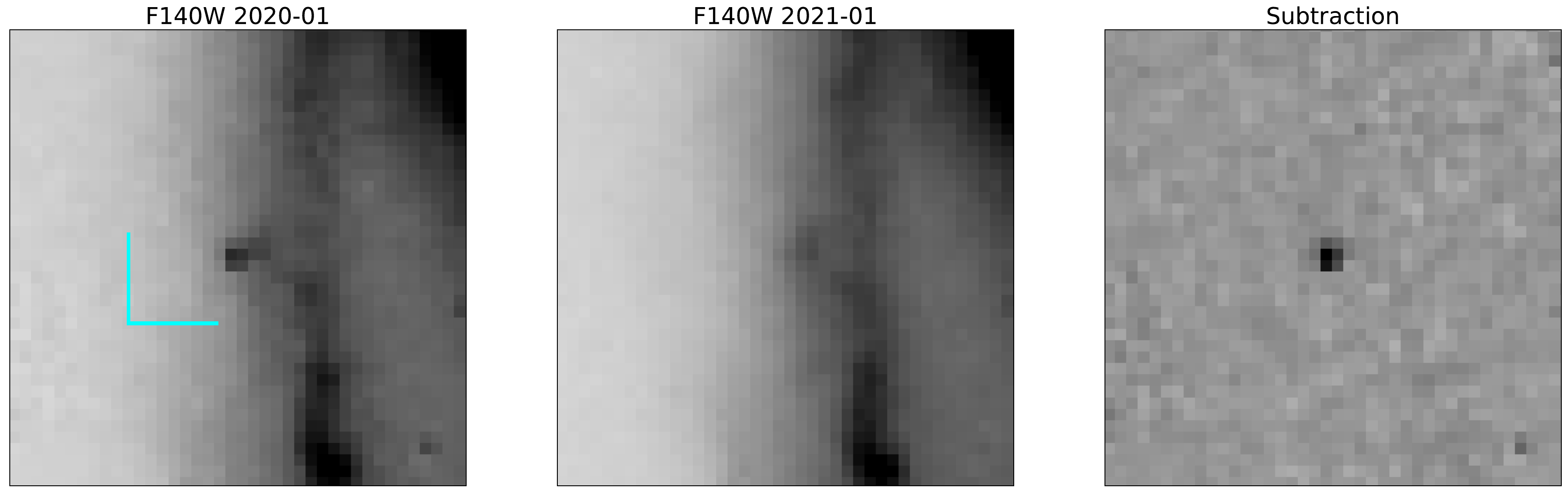}
\caption{Images of SN~2019oyw from F140W filter. The source image taken on 2020-01-12 (LEFT) was subtracted by the template image taken on 2021-01-09 (MIDDLE), resulting in the subtracted image (RIGHT) that clearly shows the transient. These images are shown in inversed-color (i.e., black = bright). Orientation is north-up and east-left. The cyan lines show the length scale of 1 arcsecond in each direction. \label{fig:f140w}}
\end{figure}

To perform the photometric measurement, we removed the galaxy light by template subtraction for filters F606W, F125W, and F140W where we used the observations on 2021-01-09 (499 days) as templates. We show in Figure \ref{fig:f140w} an example of the original source image (LEFT), template image (MIDDLE), and the result from the subtraction (RIGHT). The alignment was done by using nearby stars in a subset of the field-of view. The stellar alignments were accurate to the order of 0.1 pixel. We note some complications met when reducing some of these data. For F140W, 2020-02-11, we only combined the observations dedicated for imaging, and excluded the direct images associated with G141 grism observation. This is because the direct images associated with the grism were taken using a different part of the chip, and the distortion solutions are sufficiently inexact that the overall subtraction is better without these data, even though the total integration is somewhat shorter by excluding them (i.e., 1,200 vs. 1,900 seconds). For F140W, 2020-06-24, there was scattered light from out of the field causing the estimated upper limit to be brighter than it would have been in an uncontaminated part of the field.

\begin{figure}
\plotone{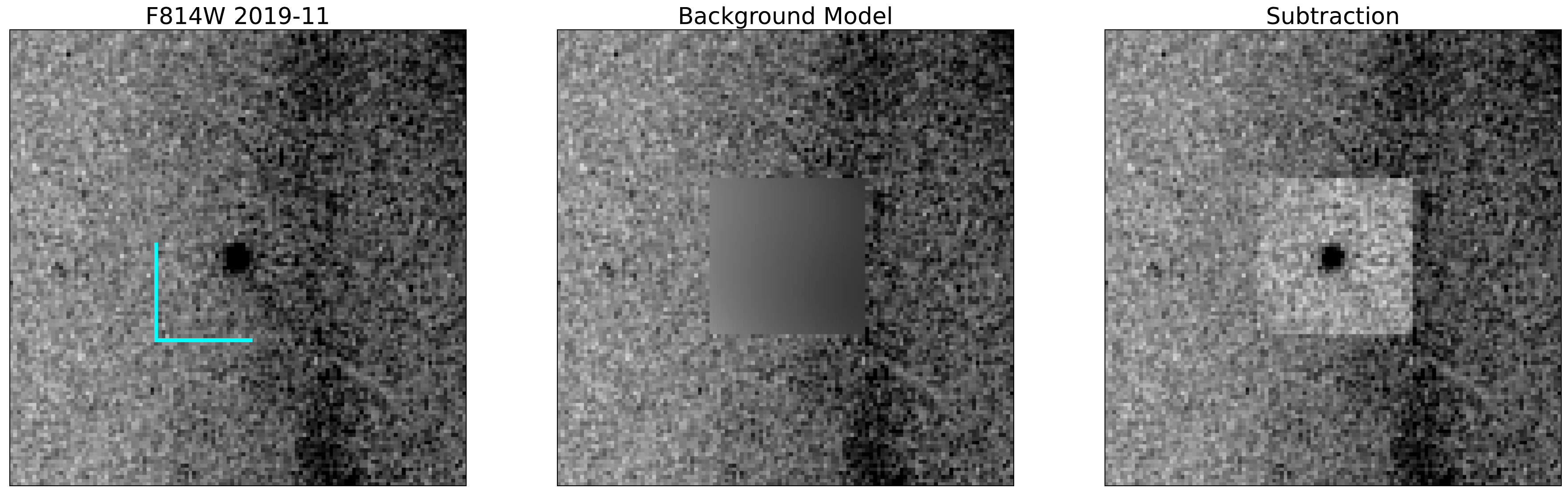}
\caption{An image of SN~2019oyw taken by $HST$/WFC3/UVIS/F814W on 2019-11-25 (LEFT). Its background model is shown in the MIDDLE image, resulting in the subtraction image on the RIGHT. These images are shown in inversed-color (i.e., black = bright). Orientation is north-up and east-left. The cyan lines show the length scale of 1 arcsecond in each direction. \label{fig:f814w}}
\end{figure}

For the other filters where we did not observe a template, we modelled the image around the source with two components: a point source as a 2D gaussian for the transient component, and the galaxy component as a smooth background modelled by a low-order polynomial. We implemented this using the \textsc{astropy} package \citep{astropy:2013,astropy:2018}. Figure \ref{fig:f814w} shows an example of the described procedure. From the source image (LEFT), the two-component model was fitted to the data given a region centered on the transient. We used the background component (MIDDLE) for the subtraction (RIGHT). 

\begin{figure}
\plotone{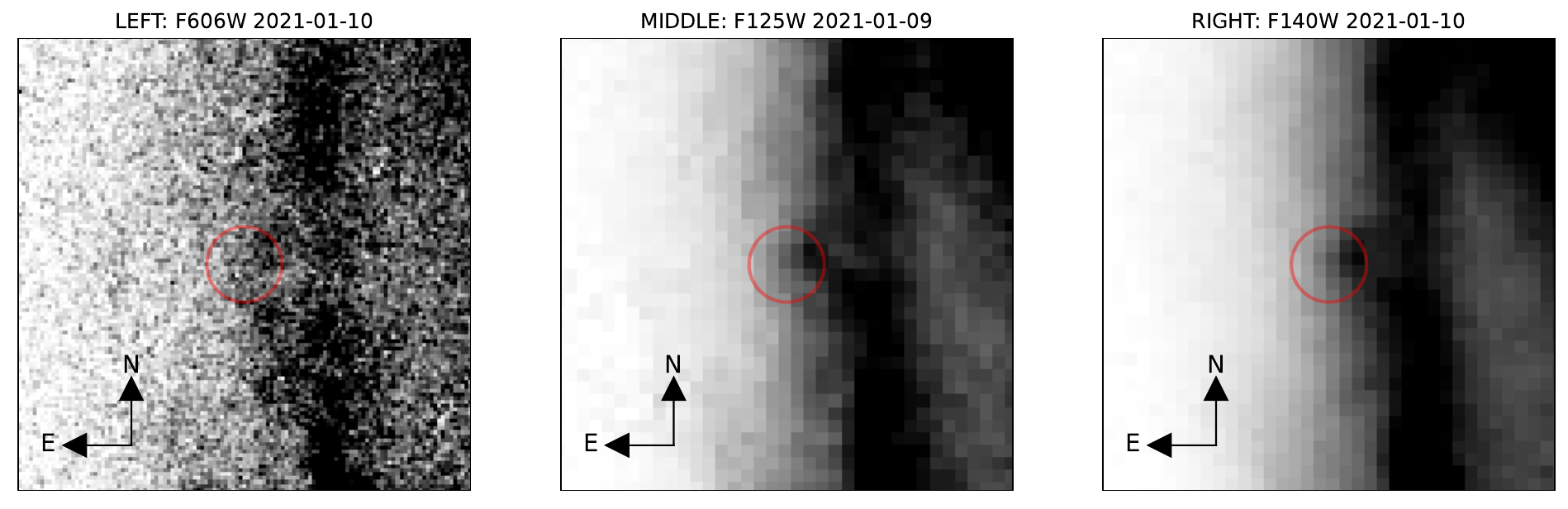}
\caption{Environment of GRB~190829A/SN~2019oyw in F606W, F125W, and F140W. Images were taken in January 2021, and are shown in inverse color, i.e., black = bright. The event was located at the center. Each red circular aperture is the size of 0.39 arcsecond in radius (i.e., 3 pixels in WFC3/IR and 9.75 pixels in WFC3/UVIS images). We observed a dust lane extending from north to south westwards from the aperture. We noted a bright clump (noticeable especially in F125W and F140W) associated with the dust lane on the western edge of the aperture. \label{fig:202101}}
\end{figure}

We note that in the template images (Figure \ref{fig:202101}) a clump, located slightly to the north and west of the SN, is noticeable in F125W and F140W, but less noticeable in F606W. Our background models with low-order polynomials cannot capture the clump. However, the method was only applied to observations from epochs 2019-09 to 2020-01, when the transient was visibly brighter than the background. To estimate the uncertainties, in order to compare with the template subtraction method we repeated the reduction process with images on 2020-01 in F606W, F125W, and F140W by applying background modelling method. We found that the subtracted images from the background models are consistently under-subtracted, when compared to the template subtraction. The difference of the photometric measurements between the two methods is about 6.5\% (F606W), 13.4\% (F125W), and 32.6\% (F140W). Since the F814W and F105W images on 2020-01 were processed by the background modelling method, the uncertainties of these photometric measurements should be approximately 10\% (if assuming a monotonic trend of the uncertainties considering the observed wavelengths).  For the 2019-10 epoch of F160W, we also estimate an error of $\sim 10\%$.

\begin{deluxetable*}{cccccccc}
\tablenum{2}
\tablecaption{Photometry of SN~2019oyw. \label{table:photometry}}
\tablewidth{0pt}
\tablehead{
\colhead{Filter} & \colhead{2019-09} & \colhead{2019-10} & \colhead{2019-11} &
\colhead{2020-01} & \colhead{2020-02} & \colhead{2020-06} & \colhead{2020-08}
}
\decimalcolnumbers
\startdata
F606W & - & - & 24.64(0.03) & 25.24(0.04) & 25.97(0.08) & $>$27.42 & $>$27.45 \\
F814W & - & - & 23.31(0.02) & 24.32(0.04) & - & - & - \\
F105W & - & - & 22.60(0.02) & 23.63(0.04) & - & - & - \\
F110W & 20.21(0.00) & 21.46(0.01) & - & - & - & - & - \\
F125W & - & - & 22.69(0.02) & 23.77(0.05) & 24.64(0.09) & - & - \\
F140W & - & - & 23.17(0.02) & 24.29(0.05) & 25.33(0.12) & $>$26.21 & $>$26.47 \\
F160W & 20.47(0.01) & 21.98(0.03) & - & - & - & - & - \\
\enddata
\tablecomments{a) Unit is in AB-magnitude. b) One-sigma uncertainties are in parentheses. c) Upper limits are 3-sigma above zero. d) These values are uncorrected for extinction. e) Epochs are shorten to only YYYY-MM format for better readability. f) Note that for F606W, F125W, and F140W we used the epoch 2021-01 as a template for host subtraction, while background models were used for the other filters (see Subsection \ref{subsec:reduction-phot}). g) Due to the lack of templates for filters F814W, F105W, F110W, and F160W, we estimate systematic uncertainties to be about 10\% on epoch 2020-01, and should be proportionately less for earlier epochs. See text.
}
\end{deluxetable*}

\begin{figure}
\plotone{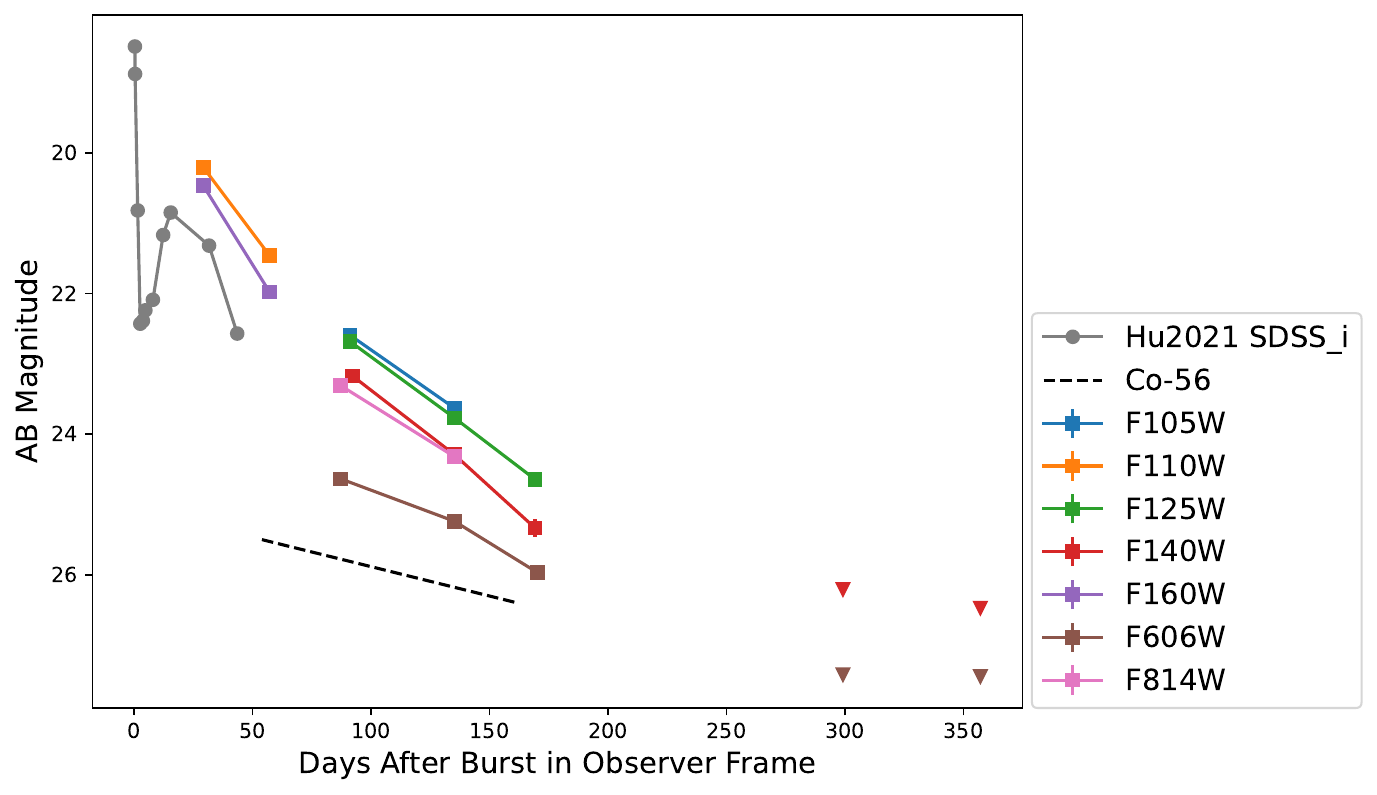}
\caption{Light curves of SN~2019oyw. From the late-time observations, data points are squares for detections, and triangles for 3-sigma upper limits. For comparison, the plot also shows observations presented in \cite{Hu2021} in SDSS's $i$-filter (gray dots), and the expected light curve's slope if powered by fully-trapped Co$^{56}$ (black dashed line). These values are uncorrected for Galactic and Extragalactic extinction. \label{fig:phot}}
\end{figure}

Aperture photometry was performed using the \textsc{Photutils} package \citep{Bradley_Photutils}.\footnote{\url{https://photutils.readthedocs.io/en/stable/}} An aperture with radius of 3 pixels was used for WFC3/IR observations (i.e., $\sim$0.39 arcsecond), and 4 pixels for UVIS (i.e., $\sim$0.16 arcsecond). Aperture correction (from finite to infinite aperture) was done following the tables given in \cite{WFC3_Instrument_Handbook}. For a non-detection, we report a 3$\sigma$ upper limit (above zero) instead. Uncertainties were propagated for the template subtraction, while we assumed negligible uncertainties from the low-order polynomial background model. Table \ref{table:photometry} summarizes the photometry in apparent AB-magnitude. Figure \ref{fig:phot} shows the light curve.

The data shows SN~2019oyw fading continuously until the last detection on 2020-02-16 (171 days after trigger in the observer's frame). In the figure, we additionally show the early $i$-band light curve from \citep{Hu2021} for comparison (in gray), which shows a consistent decline from the $i$-band peak. We note that the SDSS $i$-band is approximately comparable to the F814W of $HST$/WFC3. The decline rates from NIR bands are about 0.05 mag/day in observing frame between 2019-09 (29 days) and 2019-10 (57 days), which is slower than the optical decline rate implied by the $i$-band. Then, the decline rates turn to be slower at later times in both optical and NIR. SN~2019oyw evolved as expected for a typical GRB-SN 
by showing the late-time decline rates being consistent with being 
powered by $^{56}$Co with the $\gamma$-rays not fully trapped \citep{Nakamura2001ApJ...550..991N_98bw_lateLightCurve,Woosley2006ARA&A..44..507W_GRB-SN}.

\subsection{$HST$/WFC3 Spectroscopy} \label{subsec:reduction-spec}

The spectroscopic observations of SN~2019oyw were observed by grisms G102 (8,000 -- 11,500 \AA) and G141 (10,750 -- 17,000 \AA) with the WFC3/IR channel \citep{WFC3_Instrument_Handbook} in six epochs (see Table \ref{table:hst}). We reduced the data using the recommended grism reduction procedure \citep{WFC3_Data_Handbook} and the Python version of the \textsc{hstaxe} package.\footnote{\url{https://www.stsci.edu/scientific-community/software/axe}} In brief, the pre-processed data in FLT format can be downloaded from MAST (\url{https://archive.stsci.edu/}). With the source's location known from the associated direct images, the software can identify the source's trace, and calibrate its wavelengths in the grism images. The reduction procedure continues by performing flat-field calibration, modelling and subtracting the background around the target's trace, extracting the 1D spectrum, and calibrating the fluxes. We also used the recommended axedrizzle function \citep{Kuemmel2005} for co-adding spectra from dithered observations.

\begin{figure}[t!]
\plotone{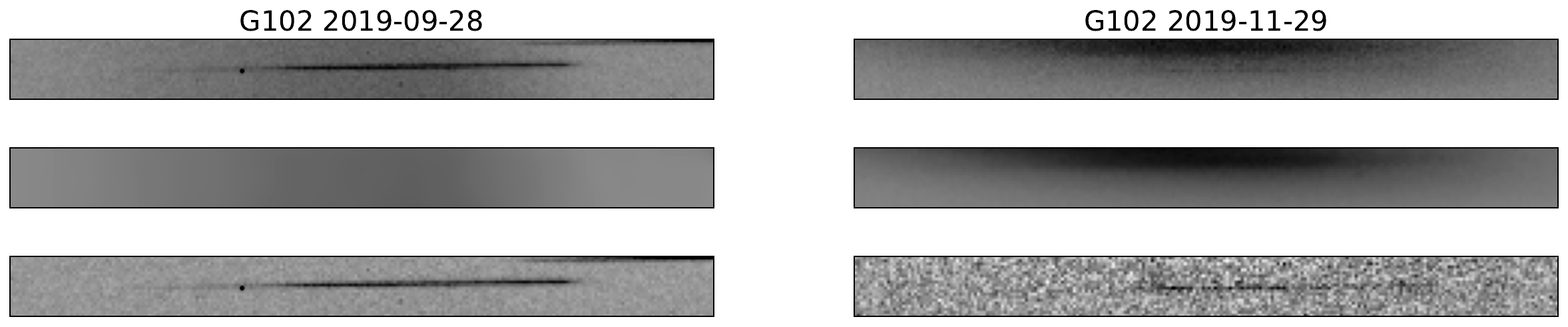}
\caption{Grism images of SN~2019oyw and their background subtractions. LEFT: G102 image on 2019-09-28 (TOP), its background model (MIDDLE), and subtraction image (BOTTOM). RIGHT: G102 image on 2019-11-29 (TOP), its background template from epoch 2021-01-09, and subtraction image (BOTTOM). These images are shown in inversed-color (i.e., black = bright). \label{fig:g102}}
\end{figure}

The templates observed on 2021-01-09 were performed with approximately the same spacecraft orientation as the observations on 2019-11-29 and 2020-01-12 to simplify and optimize the spectral template subtraction. (We note that the source was not detected on the epoch 2020-02-06). We used \textsc{drizzlepac} to generate the master template by co-adding dithered images of the template, and blotted the master template back to the orientation of each source image. Then, the blotted image was subtracted from the source image. The resulting image was then used in \textsc{hstaxe} for the reduction of the 1D spectrum. Figure \ref{fig:g102} RIGHT shows an example of this procedure: the source image (TOP), the master template (MIDDLE), and the resulting image from subtraction (BOTTOM).

For observations on 2019-09-28 and 2019-10-26, we did not observe templates for subtraction. The light contamination, mainly contributed by the SDSS galaxy, was removed by modelling. To do this, since the developer's version of the \textsc{hstaxe} (which we had access to) did not have the routine for ``local background'' estimation available, we instead used the package \textsc{hstgrism} (\url{https://pypi.org/project/hstgrism/}), that can perform this task in a similar fashion. The local background estimation (see also \url{https://github.com/spacetelescope/hstaxe}) specifies a mask over the target's trace. Then, given pixels above and below the trace mask at each wavelength (i.e., along X-axis) it performs 1D interpolation in the cross-dispersion direction (i.e., along Y-axis) assuming a low-order polynomial model. Then, a 2D Gaussian smoother is applied over the estimated background region to reduce the noises. Our best estimate specifies the trace mask of 7 pixels (in Y-axis, centered on the trace), polynomial order 2, and 3 pixels for the 2D Gaussian's sigma. Figure \ref{fig:g102} LEFT shows an example of this procedure: the source image (TOP), the background model (MIDDLE), and the resulting image from subtraction (BOTTOM).

After the galaxy-light removal, the resulting images were inputs to the \textsc{hstaxe}. We configured the software so that the \textsc{hstaxe} performed a simple box extraction and co-added using axedrizzle. We obtained the 1D spectra with uncalibrated fluxes due to the finite aperture. To calibrate to the infinite-aperture values, we adopted the aperture correction presented in \cite{Kuntschner2011}. 

\begin{figure}[t!]
\plotone{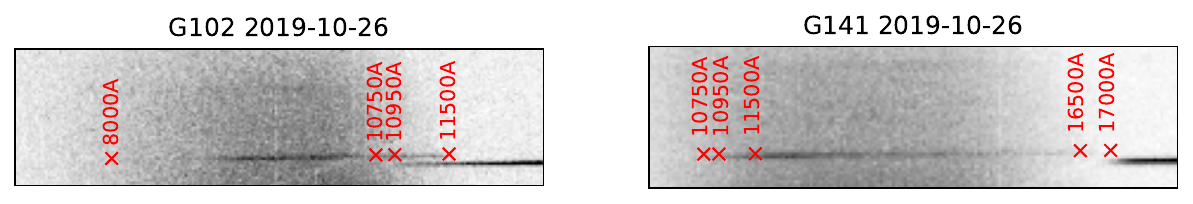}
\caption{Grism images on 2019-10-26 epoch in G102 (LEFT) and G141 (RIGHT). Red markers mark corresponding wavelengths on SN~2019oyw's traces. We note the markers 10,750 and 11,500 \AA~specifying the overlapping region between both filters. Another trace, which is visible at the bottom right of each image, was from another nearby star. The vertical separation along y-axis of the traces is 4 pixels in both filters. For G102, the star's trace contaminates starting at 10,950 \AA. For G141, SN~2019oyw's trace is truncated at the red end due to low signal-to-noise, and for this epoch we could extract meaningful information up to about 16,500 \AA, where the contamination is insignificant. We note also the truncation of the spectrum at the blue end on G102 due as well to low signal-to-noise.
\label{fig:contamination_20191026}}
\end{figure}

We also examined contamination from other nearby sources. There is a trace of a bright star locating near the red tails of the source's traces for both G102 and G141 on epochs 2019-09-28 and 2019-10-26. However, the contamination was insignificant. Specifically, on 2019-09-28 the stellar trace is well-separated from that of the GRB-SN (see on Figure \ref{fig:g102} LEFT; the star's trace is shown on the top right corner of the figure). For 2019-10-26, the contamination was also not a problem as shown in Figure \ref{fig:contamination_20191026}. For G102 (Figure \ref{fig:contamination_20191026}, LEFT), the contamination starts around 10,950 \AA. We corrected this by using the overlapping region from the G141 observation (Figure \ref{fig:contamination_20191026}, RIGHT). For G141, SN~2019oyw's trace loses its signal strengths at long wavelengths, and we could extract meaningful information to about 16,500 \AA, where the contamination was still insignificant.

\begin{figure}[t!]
\plotone{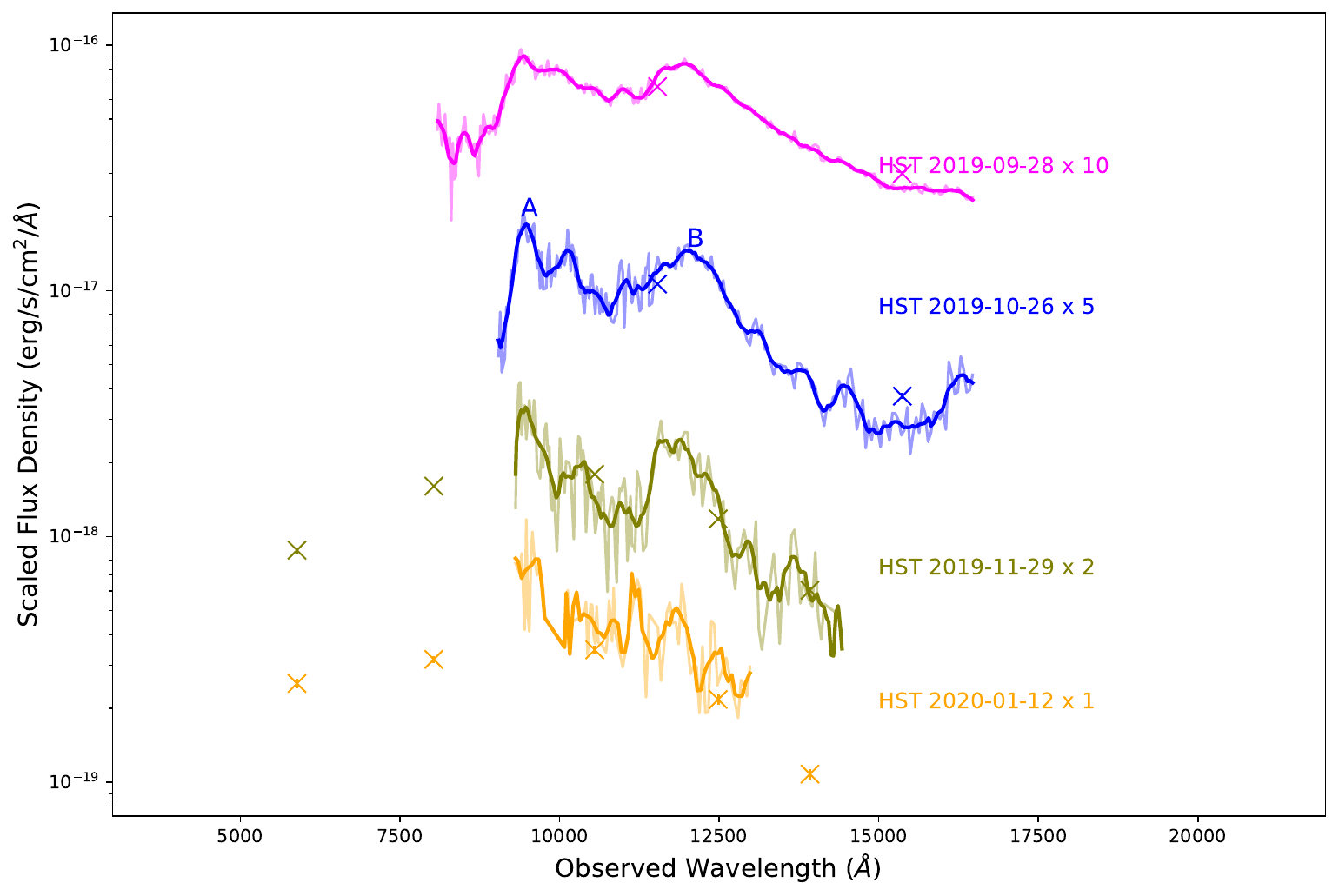}
\caption{Spectral energy distribution of SN~2019oyw from several grism spectra taken between 2019-09-28 to 2020-01-12 using {\it HST}/WFC3. The plot shows raw spectra (faint lines) and their smoothed versions (solid lines).\footnote{We applied the generalized Savitzky-Golay method \citep{Quimby2018} in smoothing the spectra. The resolution parameter was set to 60 (which is equivalent to the half width of 5,000 km/s), and the polynomial order was set to $n=2$, excepts the spectrum on 2020-01-12 using $n=0$ because of low SNR.} Each spectrum was scaled vertically for visual clarity, and the corresponding scale factor is provided. The data points shown with crosses are the photometric observations (see Table \ref{table:photometry}) shown for comparison. `A' and `B' mark two prominent features (at about 9,500 and 12,000 \AA, respectively) observed more clearly on 2019-10-26 spectrum. The observed flux densities are shown, and are not corrected for the extinction. 
\label{fig:spc}}
\end{figure}

Figure \ref{fig:spc} shows the reduced late-time spectra in the observing frame. The photometric data are also shown in the figure for flux comparison, which verifies a consistent calibration between the photometry and spectroscopy. The figure shows the declining flux density and narrowing of the spectral features with time. Note that we have cropped the three later spectra at the blue and red ends, compared to the 2019-09-26 spectrum (in magenta), due to the decrease in SNR as the source fades. Two prominent features are marked on the 2019-10-26 spectrum (in blue): `A' and `B' at about 9,500 and 12,000 \AA, respectively. These features are visible starting 2019-09-28. Feature `B', although lower SNR, was also visible on 2019-11-19 (in green). The observed spectrum on 2020-01-12 (in yellow) has such low SNR that we cannot identify any feature.

\subsection{VLT Spectroscopy} \label{subsec:data-vlt}

Additional to observations from the $HST$, \cite{Medler_190829Athesis_ljmu21959} and \Izzoinprep~provided us with their spectroscopic observations with the VLT for the analysis in this work. From \cite{Medler_190829Athesis_ljmu21959}, four epochs prior to the $HST$ observations were observed by the VLT/FOcal Reducer and low dispersion Spectrograph (FORS; \cite{Appenzeller1998Msngr..94....1A_vlt/fors}): 2019-09-02 (4 days after the trigger in observing frame), 2019-09-10 (12 days), 2019-09-15 (17 days), and 2019-09-20 (22 days). The VLT/FORS observations cover the wavelength range 3,300 -- 11,000 \AA. In Section \ref{subsec:uncertainties}, we combine this early VLT/FORS data with our later observations from $HST$ to study the evolution with time of the \CaIRfull~(see Figure \ref{fig:spc_2019oyw_vlt+hst}). \Izzoinprep~used the VLT X-shooter \citep{Vernet2011_X-shooter} to observe on 2021-09-11 and 2021-11-12. We were provided the 2D reduced spectrum (Figure \ref{fig:environment_2d}) from the combined observations. These data reveal the spectrum of the environment at the location of the transient, which we use in our discussion in Section \ref{subsec:interpret}. A full discussion of these VLT data and their implications for the GRB~190829A/SN~2019oyw will be provided in \Medlerinprep~and \Izzoinprep.

\section{Offsets of Spectral Features and Redshift Re-estimation} \label{sec:analysis}

\begin{figure}
\plotone{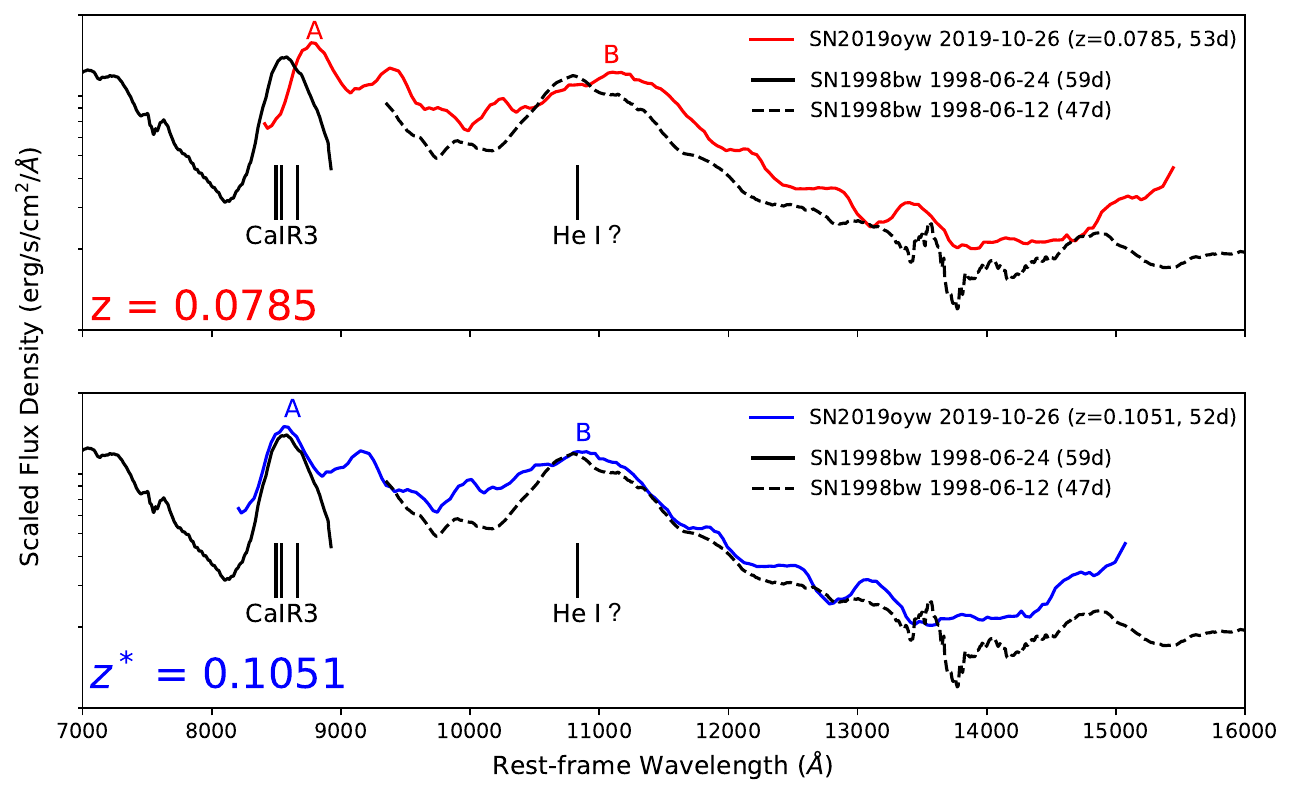}
\caption{Alignment of SN~2019oyw's late-time spectral features. Both panels show rest-frame wavelength on the x-axis, and on the y-axis the flux densities which are scaled for visual clarity. Upper Panel: Redshift $z = 0.0785$ is assumed for SN~2019oyw on epoch 2019-10-26 (red, phase 53 days) as the source. The letters `A' and `B' mark two prominent peaks in the spectrum. Spectra of both objects were extinction-corrected. SN~1998bw is chosen as the standard GRB-SN template \citep{Cano2014_ThreeGRB-SNe_2013ez_slowest,Modjaz2016_spectra_Ic_IcBL_GRB-SNe}. Epochs 1998-06-12 (black dashed line, 47 days) and 1998-06-24 (black solid line, 59 days) cover rest-frame wavelengths comparable to our observations in G102 and G141 grisms, and are comparable in phase to our observations of SN~2019oyw. The \CaIRfull~(CaIR3) and \HeIfull~are markers close to A and B, respectively. For SN~2019oyw, we adopted values $E(B-V) = 0.049, 0.757$ for Galactic \citep{Schlafly2011,Hu2021} and Extragalactic (Milky-Way model) reddening \citep{ZhangLuLu2021_GRB190829A+environment}, respectively. For SN~1998bw, we adopted $E(B-V) = 0.059$ and its host's $A_V = 0.17$ assuming Small Magellanic Cloud dust model (see \cite{Li2018_GRBhosts} and references therein). Lower Panel: Similar to the upper panel but assuming $z^* = 0.1051$ for SN~2019oyw instead (blue, phase 52 days). By comparing between the two cases, it is evident that assuming $z^*$ improves the alignment. \label{fig:0.1_vs_0.0785}}
\end{figure}

We start our analysis by comparing the late-time $HST$ spectra of SN~2019oyw with spectra of SN~1998bw \citep{Patat2001_SN1998bw}, which has served as an archetype of GRB-SNe \citep{Cano2014_ThreeGRB-SNe_2013ez_slowest,Modjaz2016_spectra_Ic_IcBL_GRB-SNe}. In Figure \ref{fig:0.1_vs_0.0785} upper panel, we show SN~2019oyw's spectrum at the epoch 2019-10-26 (the solid red line). This corresponds to a rest-frame phase of 53 days from the GRB trigger at $z = 0.0785$, the redshift of the SDSS galaxy. The letters `A' and `B' again mark the two prominent peaks at about 8,700 and 11,000 \AA~in rest-frame, respectively. The black dashed line and the black solid line show spectra of SN~1998bw on 1998-06-12, or phase 47 days, and 1998-06-24, phase, 59 days, respectively. These two spectra, obtained from  WISeREP (\url{https://www.wiserep.org/}; \cite{Yaron2012WISeREP}), are combined here to cover a rest-frame wavelength range roughly equivalent to that observed using the G102 and G141 grisms on SN~2019oyw. Black vertical lines mark the rest wavelengths of the emission line of CaIR3 (for the \CaIRfull) and He I (for the \HeIfull). 

Immediately, one notices that the features of the two transients appear offset. To better understand this issue, we experimented by assuming different redshifts for SN~2019oyw, and found an approximate redshift of $z = 0.1$.  We improved upon this using cross-correlation (see Appendix \ref{subsec:crossCorr} for details) arriving at a best estimate of  $z^* = 0.1051$. (We will discuss the errors of this estimate later, when we further discuss the time evolution of the CaIR3 peak in the following section). In Figure \ref{fig:0.1_vs_0.0785} lower panel, the redshift $z^*$ is instead assumed for SN~2019oyw (in blue solid line). Assuming $z^*$ shows better feature alignment across the {\it entire} spectrum. This implies that the features `A' agrees well with the wavelength of the  CaIR3 complex, and feature 'B' with He I;  however, as described in the Discussion (Section \ref{sec:sum}), there are good reasons to believe that elements other than He may be responsible for this feature.

\begin{figure}
\plotone{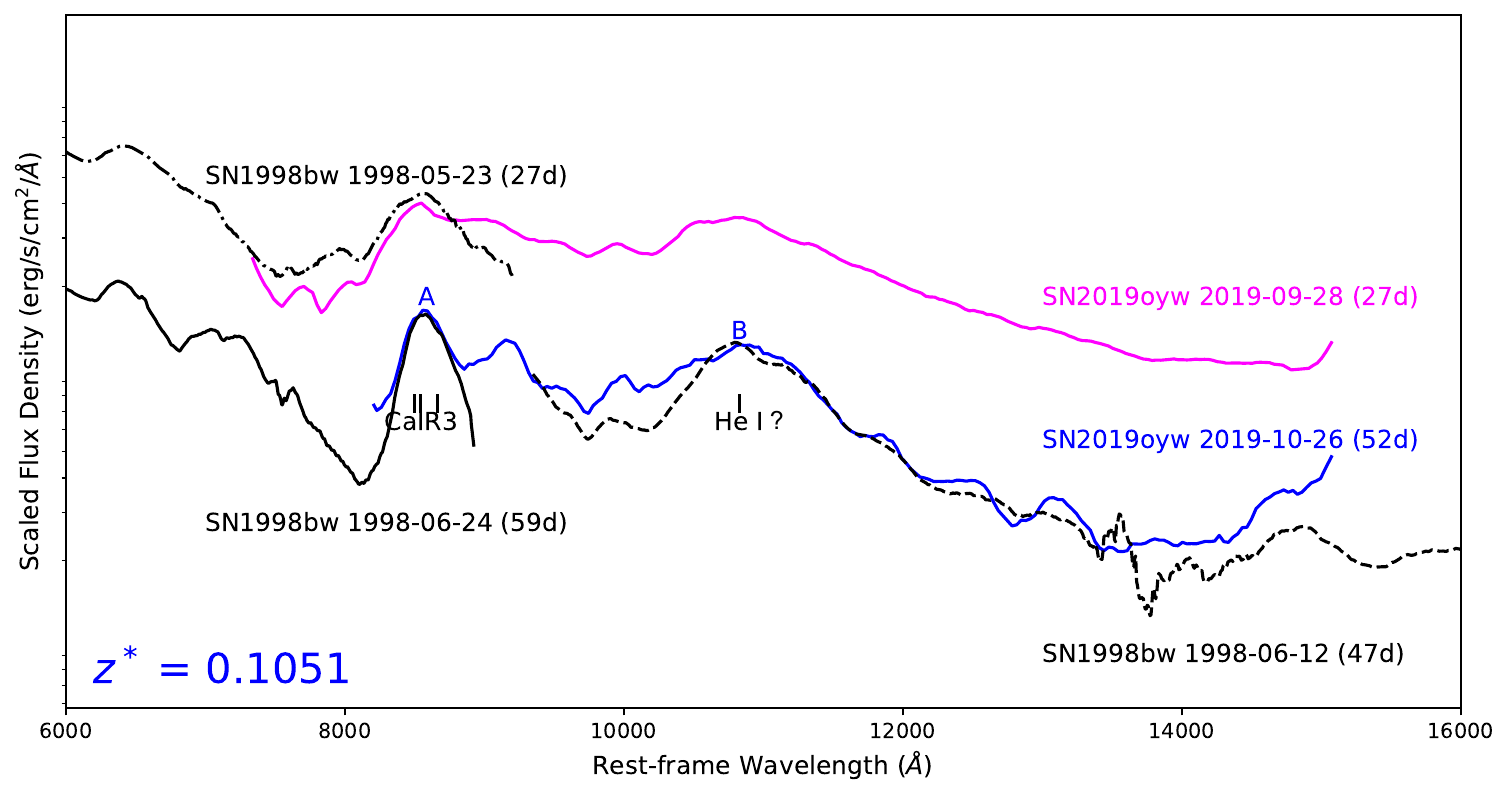}
\caption{Alignment of SN~2019oyw's late-time spectral features assuming $z^* = 0.1051$. The figure shows rest-frame wavelength on the x-axis, and on the y-axis the flux densities which are scaled for visual clarity. Spectra on epochs 2019-09-28 (magenta, 26 days), and 2019-10-26 (blue, 52 days) are shown (omitting epochs 2019-11-29 and 2020-01-12 spectra due to low signals). Spectra of SN~1998bw at similar phases are shown for comparisons. These spectra were extinction corrected. The figure supports consistency of the feature alignment assuming $z^*$. \label{fig:0.1_overview}}
\end{figure}

We build upon this in Figure \ref{fig:0.1_overview} by including the spectrum from the epoch 2019-09-28. At the redshift of $z^* = 0.1051$, this is equivalent to phase 27 days. The spectrum of SN~2019oyw (in pink solid line) is shown against SN~1998bw on 1998-05-23 (in black dotted-dashed line) which also is at phase 27 days. We repeat the epochs from $\sim$50 days in this figure to better allow comparison. The plot shows consistent alignment of SN~2019oyw's prominent feature `A' in both epochs, again assuming $z^* = 0.1051$. (SN~1998bw's spectrum at the earlier epoch does not extend sufficiently in the red to check the alignment of feature `B'). However, the alignment of both epochs strongly supports the contention that SN~2019oyw has an apparent redshift which is significantly greater than that of the large spiral on which it is superposed.

\section{Evolution of Calcium II NIR Triplet during late photospheric phase in supernovae Type Ic-BL and Ic} \label{subsec:uncertainties}

We show in the previous section that by assuming $z^* = 0.1051$, instead of $z = 0.0785$, the late-time spectral features of SN~2019oyw can be much better aligned with templates from SN~1998bw. The cross correlation we used for fine-tuning the redshift works by aligning the CaIR3 peaks of the source and the template (see Appendix \ref{subsec:crossCorr}). However, as we will discuss in this section, this feature and its peak wavelength will change over time as the expanding photosphere slows. Here, we investigate the time evolution of the CaIR3 feature and compare SN~2019oyw with samples of SNe Ic-BL and Ic.

\begin{figure}
\plotone{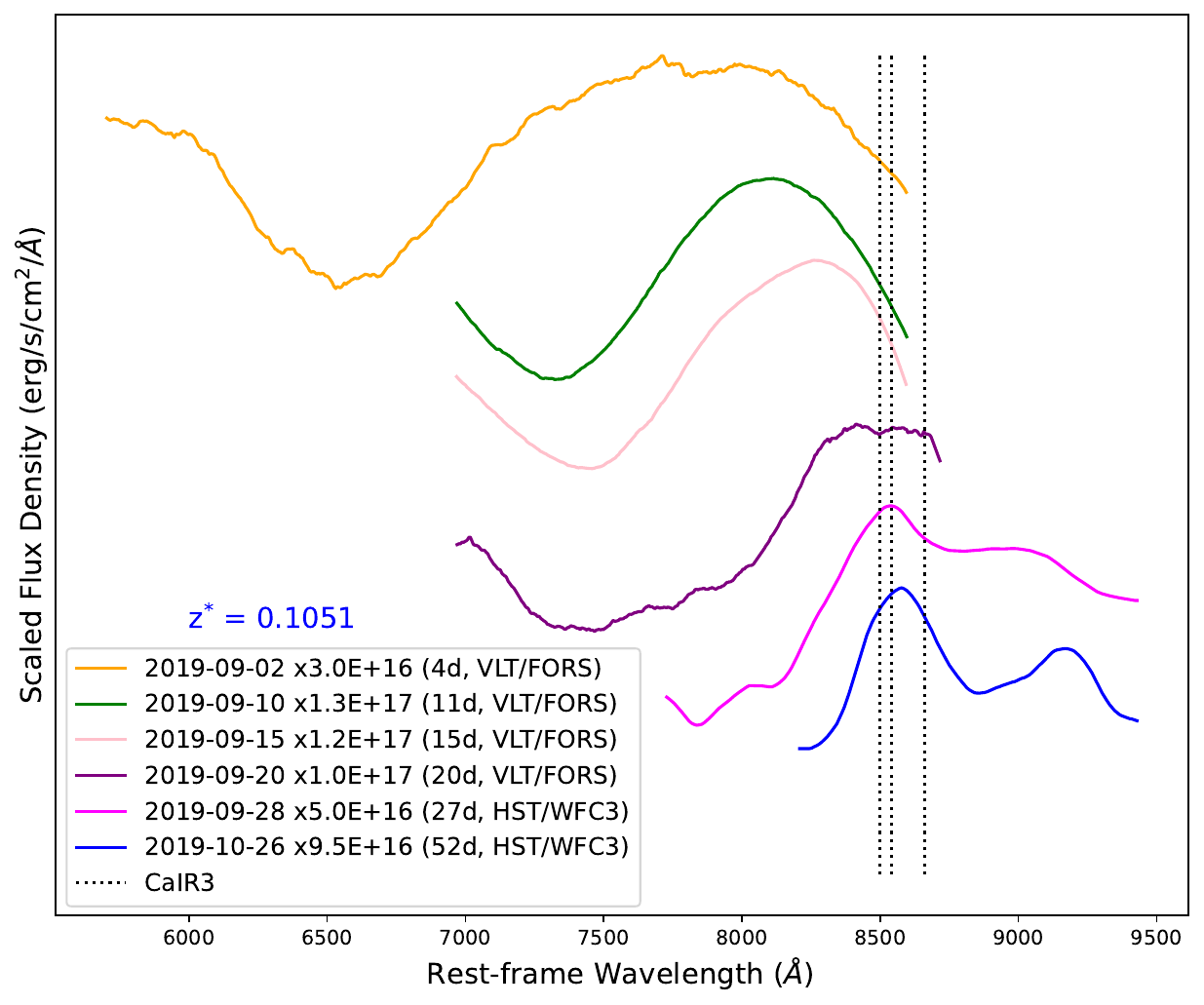}
\caption{Evolution of SN~2019oyw's CaIR3 feature. The spectra in this plot show the evolution of the CaIR3 feature from 4 days after burst to 52 days after burst (in rest-frame, assuming $z^* = 0.1051$). The spectra are scaled for legibility and do not represent the evolution of the spectral flux; the scaling factors are provided in the legend. The peak evolves from the blue back towards the restframe wavelengths of the CaIR3 interval as the expansion of the photosphere slows. As we will discuss, this is a common feature of Type Ic and Ic-BL supernovae, and understanding this time evolution is critical for accurately estimating the true redshift of SN~20190yw. These spectra are extinction corrected, and smoothed (see Subsection \ref{subsec:reduction-spec} for details) . \label{fig:spc_2019oyw_vlt+hst}}
\end{figure}

Figure \ref{fig:spc_2019oyw_vlt+hst} shows the time evolution of CaIR3 observed on SN~2019oyw's spectra from 2019-09-02 to 2019-10-26. Prior to three weeks after the explosion we use, with permission, VLT/FORS spectra from \citep{Medler_190829Athesis_ljmu21959} (see also Subsection \ref{subsec:data-vlt}). Later spectra are the \textit{HST} observations presented above. The CaIR3 feature displays a characteristic P-Cygni profile. This implies the supernova is in its photospheric phase \citep{Sim2017hsn..book..769S_SN_photosphericPhaseSpectra} throughout. We note that for a supernova the transitioning between photospheric and nebular phases is typically about 100 days after explosion \citep{Prentice2022_CaRICH}. The feature and its peak are clearly seen to migrate from blue to red with time.

\begin{figure}
\plotone{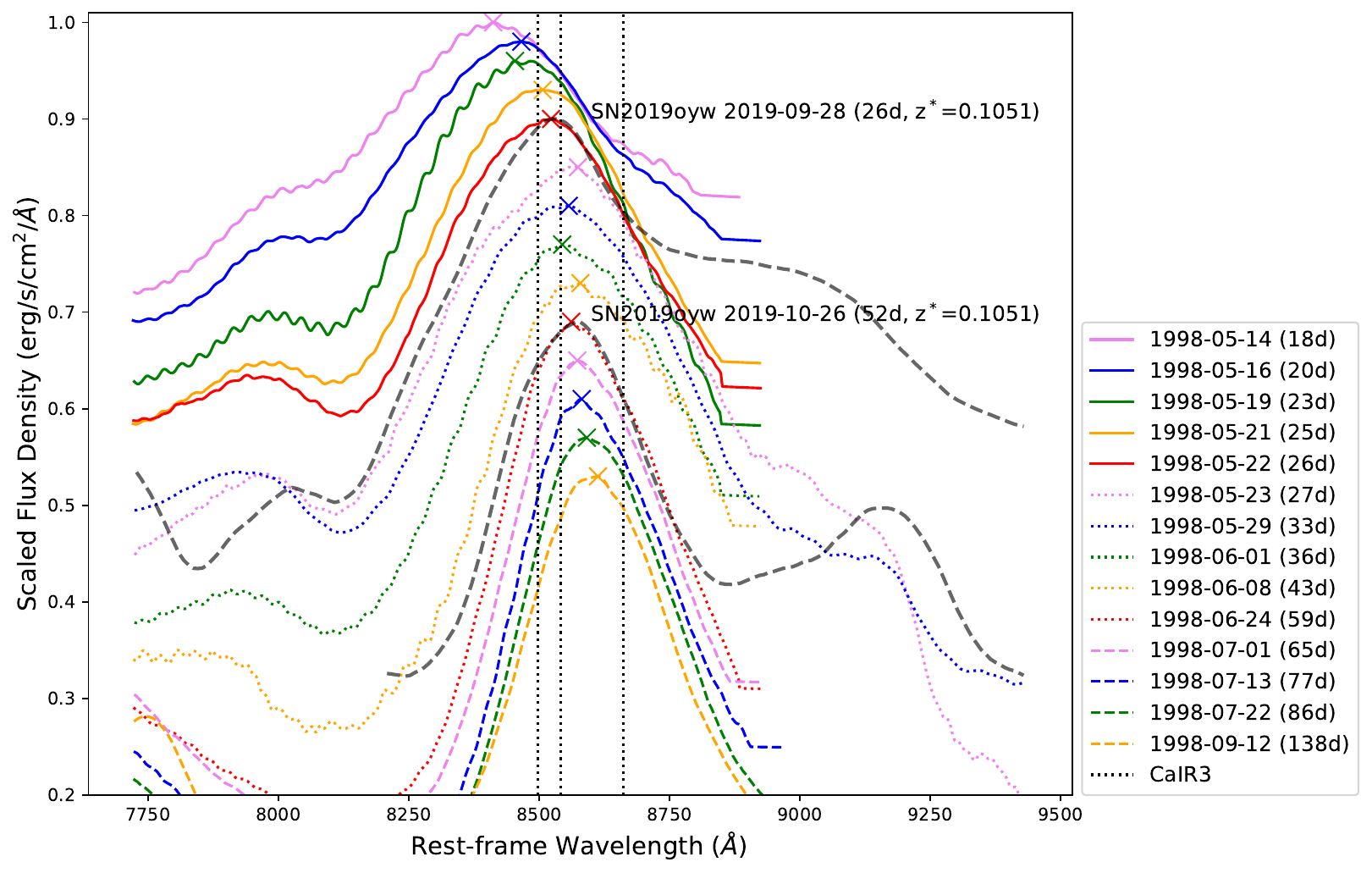}
\caption{Evolution of SN~1998bw's CaIR3 emission. The figure shows rest-frame wavelengths on x-axis, and flux densities on y-axis are scaled for visual clarity SN~1998bw's spectra from 1998-05-14 (18 days) to 1998-09-12 (138 days) are shown ordering from top to bottom by epoch \citep{Patat2001_SN1998bw}. The figure features peaks of CaIR3 (marked by crosses). The evolution shows that the peaks become redder with age. They started to be inside the constraint starting at phase 25 days, and stayed inside the constraint until at least 138 days. We discuss more about the evolution in Subsection \ref{subsec:uncertainties}. Spectra of SN~2019oyw on epochs 2019-09-28 and 2019-10-26 (black dotted lines) are also shown in the figure (assuming $z^* = 0.1051$) for comparisons with SN~1998bw at similar phases. These spectra were smoothed by the generalized Savitsky-Golay method (\cite{Quimby2018}; see Subsection \ref{subsec:reduction-spec} for details), and extinction corrected. We also note that the small oscillation (also known as ``fringing'') observed on some of SN~1998bw's spectra (e.g., 18 and 23 days) is the feature seen in the original data, not an artifact from our process.
\label{fig:SN1998bw_CaIR3}}
\end{figure}

Figure \ref{fig:SN1998bw_CaIR3} shows the time-evolution of CaIR3 features of SN~1998bw \citep{Patat2001_SN1998bw}. The CaIR3 feature of this archetypal GRB-SN evolved in a very similar fashion. For comparison, we also plot SN~2019oyw's CaIR3 feature from epochs 2019-09-28 (26 days) and 2019-10-26 (52 days) assuming $z^* = 0.1051$ in the figure along with the SN~1998bw data. The alignment of these two SNe supports the matching of the evolution of the P-Cygni profiles of CaIR3 if $z^*$ is assumed.

\begin{figure}
\plotone{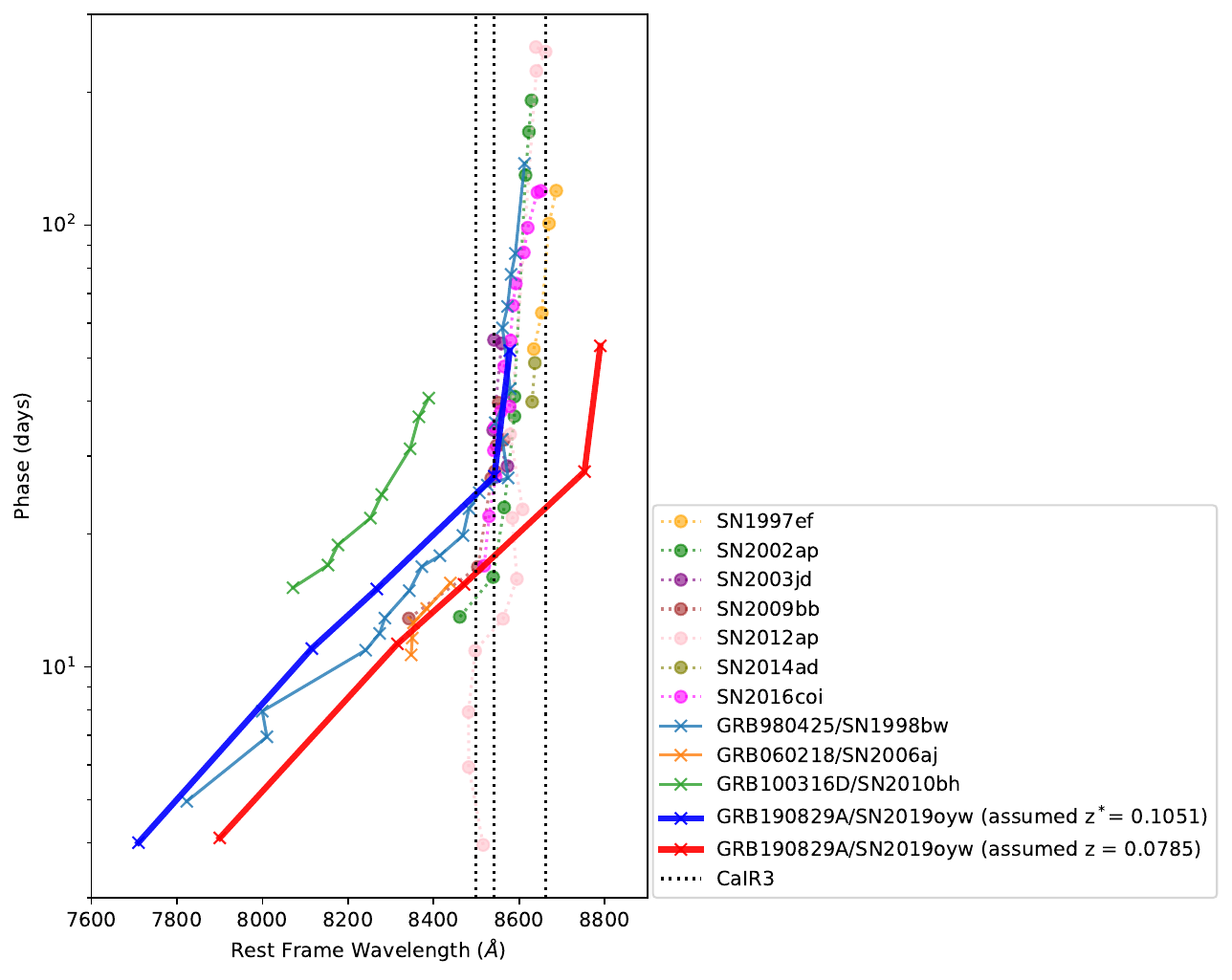}
\caption{Evolution of CaIR3 peaks from samples of SNe Ic-BL. Rest-frame wavelengths are shown on x-axis, and phases are on y-axis. The samples include SNe Ic-BL with associated GRBs (cross data points), and without GRB (dot data points). Phase is referred from the GRB trigger. The samples show the evolution of CaIR3 peaks from blue to red with ages, and the peaks are well constrained within the CaIR3 constraint if considering late-photospheric phases between 30 -- 100 days; see Subsection \ref{subsec:uncertainties} for more discussion. SN~2019oyw's CaIR3 peaks are also shown in red (assuming $z = 0.0785$) and blue (assuming $z^* = 0.1051$). We noted that early data of SN~2019oyw from VLT/FORS (see Subsection \ref{subsec:data-vlt}). The plot shows that assuming $z^* = 0.1051$ improves the feature alignment better than $z = 0.0785$.
\label{fig:CaIR3evo}}
\end{figure}

With both GRB-SNe~1998bw and 2019oyw showing similar evolution of the CaIR3 feature during the photospheric phase, we further investigate this commonality by showing the evolution of the CaIR3 peak emission in a larger sample of Type Ic-BL SNe in Figure \ref{fig:CaIR3evo}. The sample includes SNe both with and without associated GRB. (See \cite{Modjaz2016_spectra_Ic_IcBL_GRB-SNe} and references therein for further references of these objects). The figure shows the rest-frame wavelength of the evolving CaIR3 peak (in each object's rest frame) on the x-axis, and the associated phases on y-axis. SN~2019oyw's CaIR3 peak is shown in the figure in blue (assuming $z^* = 0.1051$) and red (assuming $z = 0.0785$). In the figure, we observe similar evolution of CaIR3 peaks migrating from blue towards red with age. The figure shows that the apparent peaks at early phases can be bluer than 8,498 \AA, the bluest of the CaIR3 lines. However, the evolution eventually brings the peaks back to be consistent with the CaIR3 interval during late photospheric phase. Most objects show their CaIR3 peaks residing inside the rest-frame CaIR3 wavelength range starting from 30 to 100 days. An exception is GRB-SN~2010bh which possessed significantly higher expansion velocity than the standard GRB-SN~1998bw at similar phases \citep{Bufano2012_SN2010bh,Chornock2010_SN2010bh}.

\begin{figure}
\plotone{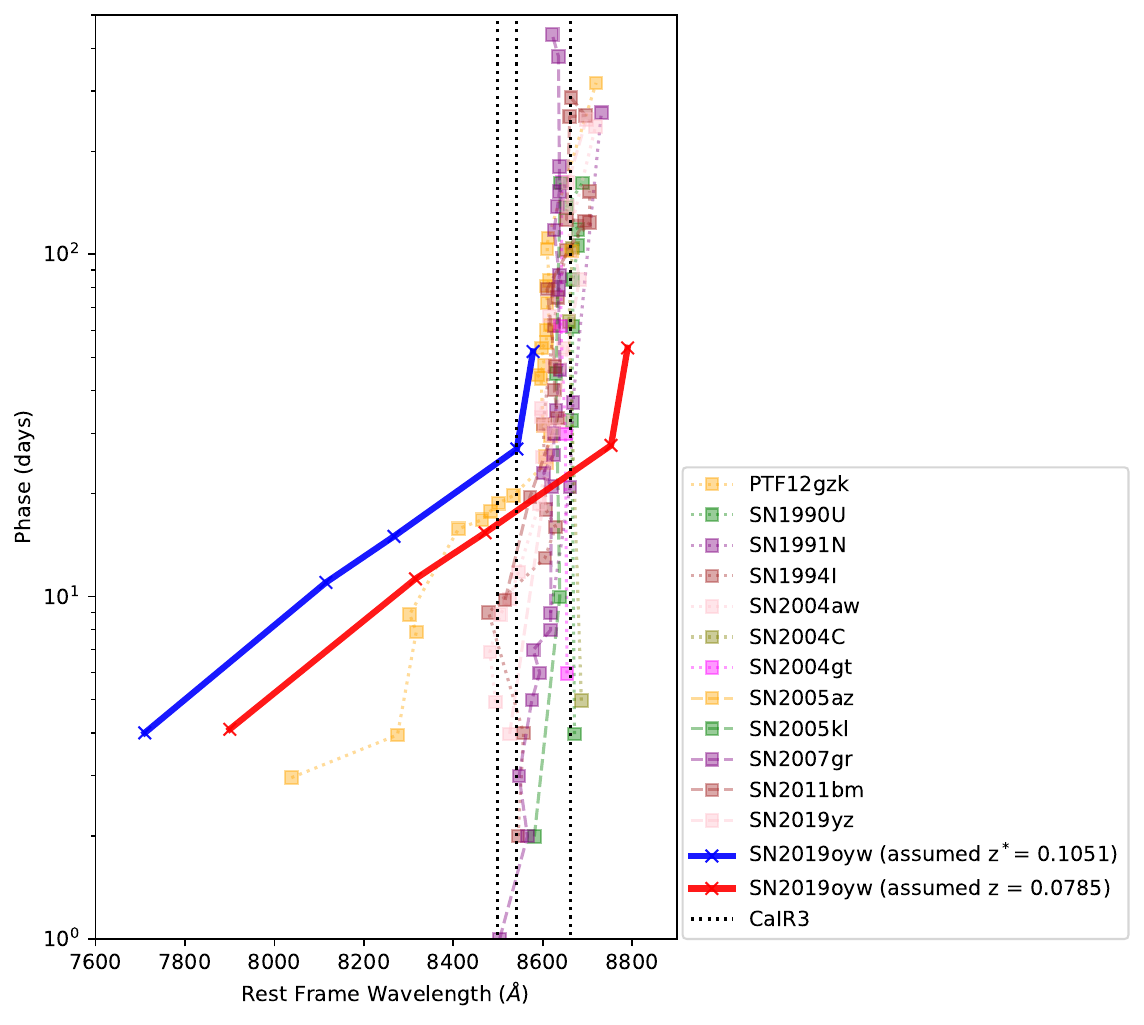}
\caption{Similar to Figure \ref{fig:CaIR3evo} but with a sample of SNe Ic (square data points) instead. We note that phase is referred from the date of discovery for these SNe without associated GRB.
\label{fig:CaIR3evo_Ic}}
\end{figure}

Similarly, Figure \ref{fig:CaIR3evo_Ic} shows the evolution of CaIR3 peaks from a sample of SNe~Ic (square data points; see also \cite{Modjaz2016_spectra_Ic_IcBL_GRB-SNe} and references therein). We note that, for a SN without GRB, which is the case for all these SNe~Ic, phase is referred from the date of discovery. The data for these transients were downloaded from WISeREP (\url{https://www.wiserep.org/}; \cite{Yaron2012WISeREP}), and we verified other relevant information such as the date of discovery and redshift from the Transient Name Server (\url{https://www.wis-tns.org/}). Like SNe Ic-BL, their CaIR3 peaks evolve to be redder with time; unlike SNe~Ic-BL, most of the SNe~Ic do not exhibit a CaIR3 peak bluer than the CaIR3 interval, due to their smaller expansion velocities. Shown among the sample, PTF~12gzk is an example of Type Ic that behaved more similarly to the Type~Ic-BL by exhibiting a bluer peak than the interval at early photospheric phase with a rapid motion to redder wavelengths, then transition to a slow motion once the peak already resides within the interval. We note that PTF~12gzk was an unusual Type~Ic exhibiting fast ejecta expansion velocities similar to Type~Ic-BL but not persistent broad lines \citep{BenAmi2012ApJ...760L..33B_PTF12gzk}.

We observe from the SNe Ic sample that their CaIR3 peaks slowly become redder while residing inside the CaIR3 interval through the late photospheric phase until about 100 days, similar to the Ic-BL sample. Beyond this, moving into the nebular phase some objects show CaIR3 peaks redder than 8,662 \AA, which may be explained by contributions from [C I] $\lambda$ 8727 and Fe II $\lambda$ 8830 \citep{Prentice2022_CaRICH}.

From these samples, we note three key observations:
\begin{enumerate}
   \item For each object, the peak of the CaIR3 evolves from blue to red with time, with this evolution being particularly rapid at early epochs when the peak is observed to be bluer than the bluest line in the triplet, i.e., 8,498~\AA.
   \item However, the CaIR3 peak evolves more slowly inside the interval, i.e., $8,498~\leq~\lambda_{\rm CaIR3}~\leq 8,662$ \AA.
   \item The peaks are well constrained within the CaIR3 interval at late photospheric phase (typically after 30 days), as the curve turns nearly vertical at this phase. 
\end{enumerate}

In Figures \ref{fig:CaIR3evo} and \ref{fig:CaIR3evo_Ic}, SN~2019oyw is displayed assuming both $z = 0.0785$ (red) and $z^* = 0.1051$ (blue). The data show the evolution of the CaIR3 peak of SN~2019oyw follows a typical path if $z^*$ is assumed. However, if $z = 0.0785$ is assumed, the CaIR3 peak of SN~2019oyw is significantly redder than any other SN in the sample, and at late times the observed spectral peak would have to somehow be transferred to another set of line or lines redder than the CaIR3 triplet. The primary suspect for these lines would be [C I] $\lambda$ 8727 and Fe II $\lambda$ 8830, which tends to appear in the nebular phase \citep{Prentice2022_CaRICH}. Therefore, we think this explanation is unlikely. 

Given the samples, the uncertainties from our cross-correlation result of $z^*$ (see Appendix \ref{subsec:crossCorr}) are dominated by two important choices we made, rather than the statistics from the cross-correlation technique itself. First, we chose the epoch 2019-10-26 for the cross-correlation as this is clearly already on the slow motion part of the CaIR3 peak's evolution which we think began around the epoch 2019-09-28. Assuming that the epoch 2019-10-26 is on the slow motion part, to behave like the other SNe in the sample, the peak wavelength must reside inside the CaIR3 interval. However, where exactly the peak should be inside the interval depends on the choice of the template. {\it Therefore, a conservative method to estimate the redshift uncertainty is to ask what redshift is required to move the CaIR3 peak on 2019-10-26 inside the CaIR3 interval, i.e., 8,498 -- 8,662 \AA. This redshift range is $0.0944~\leq~z^*~\leq~0.1156$, and we use this interval as our final estimate of $z^*$}. We note that this estimate is derived from the samples of SNe~Ic-BL and Ic presented above, not only from the SN~1998bw.

\section{Physical Scenarios} \label{subsec:interpret}

The result of redshift $z^* = 0.1051$ showing best alignment challenges our understanding of this event. GRB~190829A/SN~2019oyw superposed on the galaxy \SDSSgalaxy, and that galaxy's gas imposed absorption lines detected in the early afterglow \citep{Hu2021}. Therefore, assigning this galaxy as the host and adopting redshift $z = 0.0785$ was intuitive. However, if this was correct, the evidence presented here would imply that SN~2019oyw is a very unusual object.

If the SN~2019oyw is actually in a host at a redshift of $z = 0.0785$, the observed CaIR3 peak of SN~2019oyw correspsonds to a peculiar velocity with respect to the galaxy of $\sim$4,000~km/s in the radial direction away from the observer. Physically, this might be explained by an apparent asymmetry in the SN expansion such that line-emitting materials with high velocities away from us were dominating the observations. However, this implies that what we are seeing is dominated by light from the far side. Considering that the light from the far side should be more attenuated and obscured due to the intervening material on the closer side, how this scenario can explain the observation is uncertain. 

Another possible explanation is that the explosion occurred in a progenitor with an exceptionally high proper velocity. In a 3-body interaction \citep{Brown2015_HVSreview}, a binary (each with 3 M$_{\odot}$, with semi-major axis of the binary 0.5 AU) interacting with a massive black hole ($\sim$10$^6$ M$_{\odot}$) can disrupt one of the binary and eject the other with final velocity at infinite distance of the order of 1,000~km/s. We note that the estimated ejected velocity increases with the mass of the black hole $M$ and the mass of the binary system, $m_b$ as $M^{1/6}m_b^{1/3}$. For this event, the progenitor of GRB~190829A was a massive star (i.e., $\gtrsim$ 20 M$_{\odot}$ at zero-age main sequence; \cite{Levan2016SSRv..202...33L_GRBprogenitor}). The SDSS galaxy has nuclear emission line ratios (\Izzoinprep~and \Patricia~via personal communication), which imply the presence of an AGN, which therefore implies a supermassive black hole likely more massive than 10$^7$ M$_\odot$ \citep{Woo2002ApJ...579..530W_BHmass_AGN}. 

However, the progenitor need not have an exceptional proper motion with respect to its host galaxy, if in actuality it was in a dwarf galaxy situated behind the apparent SDSS galaxy. We note that recent studies show evidence such as very high stellar mass \citep{Gupts2022_GRB190829A_host} and the presence of AGN supporting that, if the SDSS galaxy is the host of the long GRB~190829A, it would be atypical one for a long GRB host \citep{Perley2016_grbHost_review}.

In Figure \ref{fig:202101} we show at the location of the burst the template images taken from \textit{HST}/WFC3 during January 2021 (i.e., about 500 days after the burst in the observing frame). In these images, SN~2019oyw had faded below the detection levels of the telescope. The red circle in each image shows an aperture of 0.39 arcsecond in radius centered on the supernova, using the last detection as a reference. In the vicinity of the burst location, we observed a dust lane extending from north to south on the west. We note a bright clump at the western edge of the red circle in the F125W and F140W NIR images. However, the template images do not show a clear evidence of any dwarf galaxy behind the SDSS galaxy at the burst location.

\begin{figure}[h!]
\plotone{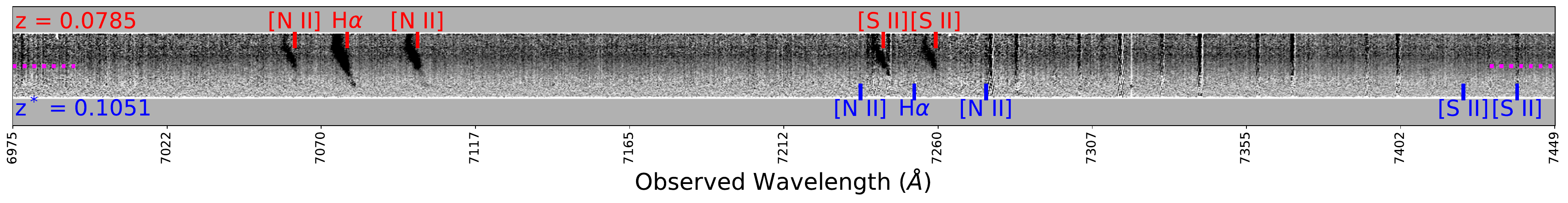}
\caption{Two-dimensional spectrum at the location of GRB~190829A/SN~2019oyw taken by X-shooter. The location of the transient was placed at the center of the slit (horizontal magenta dotted lines). X-axis is shown in observed wavelengths. The spectrum is shown in the middle, and line markers are shown at the top for assuming $z = 0.0785$ (red) and bottom for $z^* = 0.1051$ (blue). These markers include H$\alpha$ $\lambda$ 6563 \AA, [N II] $\lambda\lambda$ 6548, 6583 \AA, and [S II] $\lambda\lambda$ 6717, 6731 \AA. The image is shown in inverse color, i.e., black = bright. The observed lines are consistent with the \SDSSgalaxy~at the redshift $z = 0.0785$. See more details about this observation in \Izzoinprep.\label{fig:environment_2d}}
\end{figure}

We show in Figure \ref{fig:environment_2d} the 2D spectrum taken from X-shooter \Izzoinprep. The spectrum shows emission lines including H$\alpha$, [N~II], and [S~II] corresponding to the SDSS galaxy at $z = 0.0785$. These are marked in red. Marked in blue are the expected locations of the same set of lines assuming $z^* = 0.1051$. We note that these lines are strong emission lines typically observed in star-forming galaxies \citep{Kewley2008_MZrelation_starFormingGalaxies_ApJ...681.1183K}, which are the preferred hosts of GRB-SN events \citep{Fruchter2006_LGRB_CCSN_host,Modjaz2020_Ic_IcBL_host}. As can be seen in Figure \ref{fig:environment_2d} the spectrum reveals no clear sign of a more distant galaxy.

The non-detection does not completely reject the possibility of an existing dwarf galaxy at $z^* = 0.1051$ because it could be hidden behind the bright and dusty foreground SDSS galaxy \citep{Gupts2022_GRB190829A_host}. From the 2D spectrum, we can place an upper limit if there was a point source, such as a bright H~II region, emitting H$\alpha$ from $z^* = 0.1051$.  The estimated root mean square of the sky region around 7,253 \AA~(i.e., H$\alpha$ at $z^*$) is $3.37 \times 10^{-19}$~\flamunit~in a pixel. This is equivalent to the observing  upper limit about $6.88 \times 10^{-18}$~\flamunitintegrate~in the integrated flux assuming a point source.\footnote{For this observation from X-shooter (more details in \Izzoinprep), the wavelength dispersion (x-axis) is 2 \AA/pix with 5.7 pix/FWHM (full-width half maximum; \cite{Vernet2011_X-shooter}). We assumed a point-spread function of 0.7 arcsecond/FWHM \citep{Schonebeck2014A&A...572A..13S_xshooter_psf}, which is equivalent to 4.375 pixels/FWHM given 0.16 arcsecond/pixel in the cross-dispersion direction (y-axis).} 

We note that in this scenario the H$\alpha$ emission from $z^* = 0.1051$ passes through the SDSS foreground galaxy at $z = 0.0785$ before being observed by the telescope. Since the attenuation of the H$\alpha$ emission is mostly due to the SDSS galaxy, the dilated wavelength of the H$\alpha$ emission arriving at the SDSS galaxy would be redshifted by about $z' \approx 0.0247$ (i.e., shifted to $\sim$6,725 \AA). By assuming $A_V = 2.33$ estimated from the afterglow \citep{ZhangLuLu2021_GRB190829A+environment}, the extinction factor is $\sim$0.18. This implies the extinction-corrected upper limit $3.82 \times 10^{-17}$~\flamunitintegrate. This is equivalent to $1.08 \times 10^{39}$ erg/s (luminosity distance to $z^* = 0.1051$ is 487 Mpc). By comparing this upper limit to the H~II region luminosity function observed from nineteen nearby spiral galaxies \citep{Santoro2022A&A...658A.188S_HalphaLF}, this upper limit is at the percentile between 86$^{th}$ to 99$^{th}$ (with average 94$^{th}$). This implies that about 90 percent of these regions would be undetected considering the current upper limit.

It is perhaps interesting to ask what we would see if GRB~190829A/SN~2019oyw had gone off on a host like that of GRB~980425/SN~1998bw. The \textit{HST} has an angular resolution approximately ten times better than ground-based imaging.  Therefore, \textit{HST} imaging of the host of GRB~190829A/SN~2019oyw has roughly the same physical resolution as ground-based imaging of the host of GRB~980425/SN~1998bw, ESO 184-G82, at $z=0.0085$,  whose H~II regions were studied by \cite{Christensen2008A&A...490...45C_98bw_host_Halpha}. If we assume the missing distant galaxy at $z^* = 0.1051$ to be similar to the SN~1998bw's host, we can compare the estimated upper limit. The study found varied H$\alpha$ strengths of the H~II regions spanning from about $\sim 10^{37}$ to $10^{39}$ erg/s. All but one of these roughly two-dozen H~II regions would be $\lesssim 1\sigma$ sources in our observations were they in the hypothesized dwarf host of SN~2019oyw. Thus, to find no such source would not be too surprising.

Finally, we note that if SN~2019oyw occurred at $z^* \sim 0.1$ rather than $z=0.785$ its r-band absolute magnitude would be about 0.5 magnitude brighter than SN~1998bw instead of matching it closely \cite{Medler_190829Athesis_ljmu21959}). However, some SNe~Ic-BL both with \citep{Schulze2014A&A...566A.102S_2012bz,Toy2016_SN2013dx,Srinivasaragavan2024ApJ...960L..18S_2023pel} and without \citep{Taddia2019_SNIcBL_meta} a known GRB association have been this bright. Furthermore, \cite{Medler_190829Athesis_ljmu21959} finds that when SN~2019oyw is assumed to be at the redshift of the large $z=0.0785$ spiral, its pseudobolometric luminosity is somewhat less than that of SN~1998bw; however, we calculate that if instead it were at $z^* = 0.1051$ its pseudobolometric luminosity would agree with that of SN~1998bw to within a couple of percents.   Thus, the luminosity of the SN does not cause us to strongly prefer one distance solution over the other.

\section{Discussion and Summary} \label{sec:sum}

GRB~190829A/SN~2019oyw was observed between 2019-09-28 and 2021-01-09 using the $HST$/WFC3. Images and spectra of the transient showed a clear sign of the fading SN~2019oyw in both optical and NIR. Its late-time light curve faded with the decay rate consistent with Co$^{56}$ powering mechanism (Figure \ref{fig:phot}), typical for late-time GRB-SNe \citep{Woosley2006_GRB_progenitor} until the last detection on 2020-02-16. The spectra (Figure \ref{fig:spc}) showed broad P-Cygni profiles which narrowed and de-blended with time. In particular, two prominent peaks marked as `A' and `B' in figures were clearly visible. We attempted to identify the features by aligning the spectra against that of the standard GRB-SN~1998bw. However, we found that the spectral features were redshifted by about 4,000 km/s from the redshift of the large foreground SDSS spiral whose absorption features are observed in the afterglow spectra of GRB~190829A, $z = 0.0785$. This velocity offset can be best measured using the \CaIRfull~(CaIR3) emission, but is also seen near the \HeIfull~feature and the overall shape of the NIR spectrum (see Figure \ref{fig:0.1_vs_0.0785}). This widespread shift makes the idea that we are seeing an unusual blending of lines appear unlikely.

To better understand the behavior of the spectra, we analyzed the CaIR3 evolution in a large sample of Type Ic-BL and Ic SNe from the literature (Figures \ref{fig:CaIR3evo} and \ref{fig:CaIR3evo_Ic}, respectively). From the samples, we highlighted key observations (see Section \ref{subsec:uncertainties}). The samples show that a CaIR3 peak evolve from blue to red with time, with this evolution being particularly rapid at early photospheric phases when the peak is observed to be bluer than the bluest line in the triplet (i.e., 8,498 \AA). Inside the interval (i.e., 8,498 -- 8,662 \AA), the peak evolves more slowly, as the curve turns nearly vertical. The CaIR3 peaks are well constrained within the interval during late photospheric phases starting at about 30 days. The comparison to the samples showed no other object with the CaIR3 line as red as that of SN~2019oyw were it truly at the redshift of the SDSS galaxy. Assuming instead that SN~2019oyw behaved similarly to the SNe in the samples, we re-estimated the redshift to be $0.0944 \leq z^* \leq 0.1156$. With this new alignment, we confidently identified the feature `A' as the \CaIRfull. In contrast, we cannot be certain which spectral features are associated with the feature `B'. Without the detection of other He~I lines such as He~I~2.058~$\mu$m we cannot confirm the associate of the feature `B' with He~I rather than other features such as Mg~II \citep{Bufano2012_SN2010bh,Chornock2010_SN2010bh,Izzo2019_2017iuk,Lucy1991-HeI-1991ApJ...383..308L,Mazzali1998-HeI-1998MNRAS.295..428M,Patat2001_SN1998bw,Taubenberger2006-2004aw-HeI-2006MNRAS.371.1459T}.

As GRB~190829A/SN~2019oyw is superposed on the galaxy \SDSSgalaxy~whose gas was seen in Ca II H/K absorption in the early afterglow, assigning this galaxy as the host was intuitive. However, our surprising result challenges this picture. If the SDSS galaxy is its actual host, the evidence implies that SN~2019oyw was very unusual. The redshifted features offset by 4,000 km/s could be a consequence from an observing asymmetry such that the line-emitting materials from the far side (i.e., moving away from observers) were dominating the observations. It is far from obvious about how one gets the dominance of the far side. Another possible explanation (see Section \ref{subsec:interpret}) involves a progenitor with high velocity such as one that would be produced by a 3-body interaction of a tight and massive binary and a supermassive black hole \citep{Brown2015_HVSreview}. We note that the presence of an AGN was observed in the SDSS galaxy (\Izzoinprep~and \Patricia~via personal communication), which implies the presence of a supermassive black hole likely more massive than 10$^7$ M$_\odot$ \citep{Woo2002ApJ...579..530W_BHmass_AGN}. 

Instead, if we placed the explosion further behind the SDSS galaxy on a smaller dwarf galaxy at $z^*$, the observed discrepancies could be straightforwardly resolved. In this scenario, the spectral features of SN~2019oyw evolved similarly to other SNe Ic-BL and Ic in the comparison samples. Given that the SDSS galaxy is bright and dusty, especially at the location of the event, searching for a more distant dwarf host would be challenging. The available images from $HST$ and the spectra from the X-shooter show no sign of such a system. We note that these observations were not designed to see beyond the foreground SDSS galaxy, and were done before we knew about this surprising evidence. We estimate the upper limit of the H$\alpha$ emission from a bright H~II region in a background galaxy at $z^*$, and show that our non-detection does not by any means rule out the  presence of such a host . We encourage further investigation with a design that can overcome observing challenges and see beyond the bright and dusty SDSS galaxy, such as a radio search for H~I emission from the potential dwarf host.

In summary, we present a surprising result on the origins of the nearby GRB~190829A. Our story is similar to GRB~020819B \citep{Perley2017MNRAS.465L..89P_GRB020819B_wrongHost} and GRB~130702A \citep{Kelly2013_GRB130702A_dwarf_satellite_host} in that there is new evidence supporting a possible revision of the assigned host of the event. The possible need for such a revision is not a total surprise as several studies had already pointed out that the SDSS galaxy is quite atypical for a long-duration GRB host. In particular, the SDSS galaxy was estimated to have stellar mass on the order of 10$^{12}$ M$_\odot$ \citep{Gupts2022_GRB190829A_host}, very massive for a long GRB host. There is also evidence indicating the presence of AGN from the galaxy, which is atypical for a long GRB host. However, if the explosion was behind the foreground spiral, GRB~190829A/SN~2019oyw could be one among the typical cases like GRB~980425/SN~1998bw \citep{Galama1998_GRB-SN98bw}.

Finally, the work here provided insights into the time evolution of GRB-SNe and a potential method for directly determining the redshift of a GRB-SN. We demonstrated in Section \ref{subsec:uncertainties} that using the CaIR3 feature has advantages due to the strong signal strength, stability and slow evolution during late photospheric phase from phase about 30 -- 100 days. The accuracy of this method is of the order of 1,000 km/s. We encourage further studies of the use of CaIR3 for the redshift estimation. In contrast to other existing methods \citep{Li2023FrASS..1024317L_measureZ_GRB}, this method allows us to estimate the redshift of the explosion using the lines of the associated supernova. Furthermore, since CaIR3 is a common feature typically observed in a supernova regardless of its class \citep{Marion2009_NIRcatalogueSNeIa,Cano2014_ThreeGRB-SNe_2013ez_slowest,Gutierrez2017,Modjaz2016_spectra_Ic_IcBL_GRB-SNe,Prentice2022_CaRICH,Silverman2015,vanRossum2012_CaIR3_in_SNIa}, the use of CaIR3 for redshift estimation could benefit not only the cases of GRB-SNs, but also potentially other calcium-rich transients.

\begin{acknowledgments}

This research is based on observations made with NASA/ESA Hubble Space Telescope obtained from the Space Telescope Science Institute, which is operated by the Association of Universities for Research in Astronomy, Inc., under NASA contract NAS 5–26555. The {\it HST} observations presented here were taken under  General Observer(GO) Programs 15089, 15510, 16042, and 16320.   Support for K. Bhirombhakdi was provided through grants associated with GO Programs 15510, 16042 and 16320 from the STScI under NASA contract NAS5-26555.

This research made use of Photutils, an Astropy package for
detection and photometry of astronomical sources (Bradley et al.
2021).

This work made use of the data products generated by the Modjaz group (formerly NYU SN group), and 
released under DOI:10.5281/zenodo.58766 \citep{https://doi.org/10.5281/zenodo.58766}, 
available at \url{https://github.com/nyusngroup/SESNtemple/}.

KB thanks Dr. Robert Quimby for sharing his code that performs the generalized Savitzky-Golay method \citep{Quimby2018}.

S. B. acknowledge support from the PRIN-INAF 2022 project “Shedding light on the nature of gap transients: from the observations to the model" 

This project was conducted during the pandemic of COVID-19. We would like to thank every individual both inside and outside the scientific communities, for all the efforts during this hard time. And, we would like to offer our deepest and most sincere condolences for every loss.

\end{acknowledgments}

%% For this sample we use BibTeX plus aasjournals.bst to generate the
%% the bibliography. The sample631.bib file was populated from ADS. To
%% get the citations to show in the compiled file do the following:
%%
%% pdflatex sample631.tex
%% bibtext sample631
%% pdflatex sample631.tex
%% pdflatex sample631.tex
\appendix
\section{Using Cross Correlation to Determine Redshift of SN~2019oyw} \label{subsec:crossCorr}

\begin{figure}
\plotone{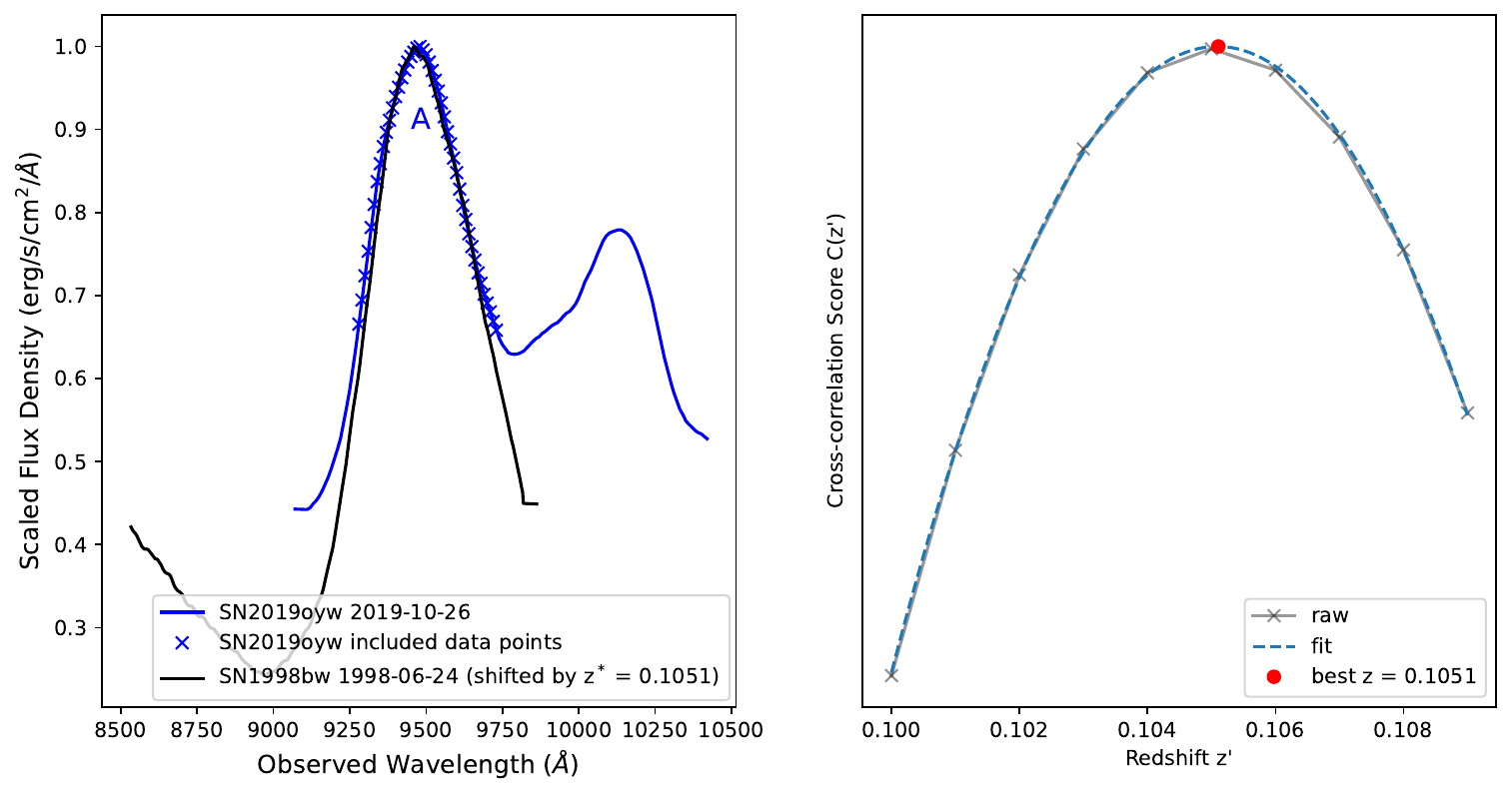}
\caption{Cross correlation matching CaIR3 peaks. LEFT: The cross correlation technique was performed using SN~2019oyw on epoch 2019-10-26 (source, blue) and SN~1998bw on epoch 1998-06-24 (template, black). For the template, this peak was identified as CaIR3 \citep{Patat2001_SN1998bw}. For the source, only included data points around the peak (blue crosses) were used in the computation. For more details, see Appendix \ref{subsec:crossCorr}. RIGHT: The optimal solution $z^*$ from the cross-correlation (red dot) is estimated by the second-order polynomial fit to the cross-correlation score curve. \label{fig:crosscorr_bestfit}}
\end{figure}

In this section, we present our use of cross-correlation \citep{Blondin2007_crosscorr} to estimate the redshift of SN~2019oyw. Our aim was to perform the cross correlation around the CaIR3 emission of SN~1998bw and the feature `A' of SN~2019oyw as shown in Figures \ref{fig:0.1_vs_0.0785} and \ref{fig:0.1_overview}. The epoch 2019-10-26 was chosen for the computation to limit the complication from line blending, as the feature is relatively narrow and high SNR at this epoch. The epoch 1998-06-24 of SN~1998bw \citep{Patat2001_SN1998bw}, which is at a comparable phase, was chosen as the GRB-SN template \citep{Cano2014_ThreeGRB-SNe_2013ez_slowest,Modjaz2016_spectra_Ic_IcBL_GRB-SNe}. 

We implemented our code to perform the cross correlation following \cite{Blondin2007_crosscorr}, except that the we did a direct cross-correlation on the data, rather than transforming to Fourier space. Both source and template spectra were extinction corrected. Following \cite{Blondin2007_crosscorr} both spectra were smoothed prior to cross-correlation.  We used the generalized Savitsky-Golay method (\cite{Quimby2018} using the same parameters as for our earlier spectral reduction (see Section \ref{subsec:reduction-spec} for the parameters). The template spectrum was normalized with respect to its CaIR3 peak, and the source to its feature `A'. Due to limitation on the template spectrum longer than 8,800 \AA~(rest-frame) and on the source shorter than 9,100 \AA~(observed frame), we chose to include data points surrounding the peak of the feature `A' (approximately 9,300 to 9,700 \AA~in the observed frame, or 8,400 to 8,800 \AA~in the rest-frame assuming $z^* = 0.1051$ as shown in Figure \ref{fig:crosscorr_bestfit} LEFT).

We performed the cross correlation by assuming the redshift $z' \in [0.100,0.109]$ with a step of 0.001. The cross-correlation score $C(z = z')$ showed a strictly concave profile, which was fit by a second-order polynomial to locate the optimal solution. As shown in Figure \ref{fig:crosscorr_bestfit}, the optimal solution is $z^* = 0.1051 \pm 0.0009$ as shown in Figure \ref{fig:crosscorr_bestfit} RIGHT.

\bibliography{bib}{}
\bibliographystyle{aasjournal}

%% This command is needed to show the entire author+affiliation list when
%% the collaboration and author truncation commands are used.  It has to
%% go at the end of the manuscript.
%\allauthors

%% Include this line if you are using the \added, \replaced, \deleted
%% commands to see a summary list of all changes at the end of the article.
%\listofchanges

\end{document}